\documentclass[12pt]{iopart}

\usepackage{array,amsfonts}
\usepackage{cite,wrapfig,color}
\usepackage{graphicx}

\newcolumntype{x}[1]{%
>{\centering\hspace{0pt}}m{#1}}%
\newcolumntype{w}[1]{%
>{\raggedright\hspace{0pt}}m{#1}}%
\newcolumntype{z}[1]{%
>{\raggedleft\hspace{0pt}}m{#1}}%

\newcommand{\fud}[2]{{^{#1}{}_{#2}}}
\newcommand{\fdu}[2]{{_{#1}^{\phantom{#1}#2}}}

\newcommand{\DL}{{D}}

\newcommand{\DO}{{D_{\Omega}}}

\newcommand{\refp}[2]{{{}^{#1}\!#2}}

\newcommand{\wiM}{{\underline{M}}}

\newcommand{\ga}{{\ensuremath{{\alpha}}}}
\newcommand{\gb}{{\ensuremath{{\beta}}}}

\newcommand{\la}{{\ensuremath{\mathsf{a}}}}
\newcommand{\lb}{{\ensuremath{\mathsf{b}}}}

\newcommand{\lm}{{\ensuremath{\mathsf{m}}}}

\newcommand{\pl}{\partial}

\newcommand{\ddd}[1]{{\frac{\partial}{\partial #1}}}

\newcommand{\be}{\begin{equation}}
\newcommand{\ee}{\end{equation}}
\newcommand{\bee}{\begin{eqnarray}}
\newcommand{\eee}{\end{eqnarray}}

\newcommand{\WeylD}[3]{\ensuremath{\left[\frac{#1}{#2}{#3}\right]}}

\newcommand{\bep}{\begin{picture}}
\newcommand{\eep}{\end{picture}}

\newcounter{YoungHeight}\newcounter{YoungWidth}

\newcounter{Mul1}\newcounter{Mul2}\newcounter{Mul3}\newcounter{Mul4}
\newcounter{A1}\newcounter{A2}
\newcounter{B3}

\newcounter{T0}\newcounter{T1}

\newcounter{R0}

\newlength{\txtHShift}

\newlength{\txtWidth}

\newcommand{\HalfLength}[2]{\setcounter{Mul1}{#1}\setcounter{Mul2}{#1}\addtocounter{Mul1}{\value{Mul2}}\addtocounter{Mul1}{\value{Mul2}}%
\addtocounter{Mul1}{\value{Mul2}}\addtocounter{Mul1}{\value{Mul2}}\setcounter{#2}{\value{Mul1}}}

\newcommand{\Add}[3]{\setcounter{#1}{#2}\addtocounter{#1}{#3}}

\newcommand{\Length}[1]{#10}
\newcommand{\YoungScale}{}

\newcommand{\shiftedText}[2]{{\hspace{#1}{#2}}}
\newcommand{\calcHShift}[1]{\settowidth{\txtWidth}{#1}\setlength{\txtHShift}{-0.5\txtWidth}}

\newcommand{\TextTop}[3]{{\calcHShift{#1}\HalfLength{#2}{T0}\Add{T1}{\Length{#3}}{-9}\put(\value{T0},\value{T1}){\shiftedText{\txtHShift}{#1}}}}

\newcommand{\BlockA}[2]{{\YoungScale\bep(\Length{#1},\Length{#2}){\Add{A1}{#1}{1}\Add{A2}{#2}{1}}%
\multiput(0,0)(10,0){\value{A1}}{\line(0,1){\Length{#2}}}\multiput(0,0)(0,10){\value{A2}}{\line(1,0){\Length{#1}}}%
\setcounter{YoungHeight}{\Length{#2}}\setcounter{YoungWidth}{\Length{#1}}\eep}}

\newcommand{\RectT}[3]{\bep(\Length{#1},\Length{#2})\put(0,0){\line(1,0){\Length{#1}}}\put(0,0){\line(0,1){\Length{#2}}}%
\put(\Length{#1},\Length{#2}){\line(-1,0){\Length{#1}}}\put(\Length{#1},\Length{#2}){\line(0,-1){\Length{#2}}}#3{#1}{#2}\eep}

\newcommand{\RectBRow}[4]{{\bep(\Length{#1},20)\put(0,0){\RectT{#2}{1}{\TextTop{#4}}}%
\put(0,10){\RectT{#1}{1}{\TextTop{#3}}}\eep}}

\newcommand{\YoungB}{\BlockA{2}{1}}

\newcommand{\YoungAA}{\BlockA{1}{2}}

\newcommand{\YoungBB}{\BlockA{2}{2}}

\newcommand{\YoungCC}{\BlockA{3}{2}}

\newcommand{\YoungAAAA}{\BlockA{1}{4}}

\begin{document}

\title{Towards higher-spin holography in ambient space of any dimension}

\author{V E Didenko$^1$ and E D Skvortsov$^{1,2}$}

\address{$^1$Lebedev Institute of Physics, Moscow, Russia}
\address{$^2$Albert Einstein Institute, Potsdam, Germany}

\ead{didenko@lpi.ru, skvortsov@lpi.ru}
\begin{abstract}
We derive the propagators for higher-spin master fields in anti-de Sitter space of arbitrary dimension. A method is developed to construct the propagators directly without solving any differential equations.
The use of the ambient space, where AdS is represented as a hyperboloid and
its conformal boundary as a projective light-cone, simplifies the approach and makes a direct contact between
boundary-to-bulk propagators and two-point functions of conserved currents.
\end{abstract}

 \vspace{2pc} \noindent{\it Keywords}: higher-spin
gauge theory, AdS/CFT correspondence

\maketitle
\section{Introduction}
A canonical playground for the AdS/CFT correspondence
\cite{Maldacena:1997re, Gubser:1998bc, Witten:1998qj},
$\mathcal{N}=4$ SYM vs. superstring theory on $AdS_5\times S^5$ is
quite complicated. One wishes to have a simpler model that
captures all essential features of the AdS/CFT paradigm.
Retrospectively, it would have been more natural to look for the
AdS duals of the simplest CFT's, the free ones, rather than
strongly coupled. In the OPE of two free conformal fields one
finds an infinite set of conserved currents of increasing tensor
rank, suggesting the AdS dual be a theory of gauge, and therefore
massless, fields of all spins. Such theories, called higher-spin
theories do exist in arbitrary dimension, \cite{Vasiliev:1990en,
Vasiliev:1997dq, Vasiliev:1997ak, Prokushkin:1998vn,
Prokushkin:1998bq, Vasiliev:2003ev}, see \cite{Vasiliev:1995dn,
Vasiliev:1999ba,Bekaert:2005vh} for reviews. The conjectures that
relate them to various CFT's, which are not always free, have been
proposed in \cite{Sundborg:2000wp, Sezgin:2002rt,
Klebanov:2002ja,Chang:2012kt}, see also \cite{Konstein:2000bi, Leonhardt:2002sn, Sezgin:2003pt,
Leigh:2003gk, Diaz:2006nm, Maldacena:2011jn, Bekaert:2012ux}.

In addition to being dual to 'almost' free CFT's, the simplest
higher-spin theories have a nondegenerate spectrum of states and
they are dual to vector models, which do not have long trace
operators. Therefore higher-spin theories provide a promising
model for AdS/CFT, in which one can expect to prove everything.
However, it was not until the main breakthroughs
\cite{Giombi:2009wh, Giombi:2010vg} in $AdS_4/CFT^3$ and
\cite{Campoleoni:2010zq, Henneaux:2010xg} in $AdS_3/CFT^2$, which
is a special case, that the topic attracted considerable
attention.

One of the goals of the present research is to pave the way for a
search for the $AdS/CFT$ dual to d-dimensional bosonic higher-spin
(HS) field theory \cite{Vasiliev:2003ev}. The $AdS/CFT$ analysis
in terms of the correlation functions requires among other things
explicit form of the bulk-to-boundary propagators. We develop a
method that allows us to find boundary-to-bulk propagators for all
higher-spin fields without solving any differential equations at
all.

HS gauge theory has its dynamical content totally encoded in
terms of some {\it master} fields. These are $W$ gauge connection
one-form and $B$ higher-spin curvature zero-form. Physical
information can be equivalently extracted from either of two
fields. Pretty much as in the case of gravity where the degrees of
freedom reside in either the metric (gauge field) or the Riemann
tensor (curvature field). At nonlinear level, however, the
perturbative sector of HS connections is way more involved as
compared with curvature $B$-sector. Practically, it makes the
HS-curvature analysis sometimes more preferable to the
HS-connections one. Particularly, the AdS/CFT correspondence test
carried out in \cite{Giombi:2009wh, Giombi:2010vg} heavily rests
upon the $B$ boundary-to-bulk propagator calculation.

The aim of
this paper is to explicitly derive $W$- and $B$-propagators for all spins
in arbitrary dimension. The straightforward approach
based on solving the equations of motion encounters formidable
technical problems though. It calls for some more refined methods
to push the matter through. One of the results of our paper is the
method that effectively allows us to build $B$-propagators by
purely algebraic means. The method that we call the
self-similarity virtually represents some motley combination of
three ingredients. These are the ambient, the unfolding and the
star-product. In its essence, it makes it possible to construct a
generating function out of the spin $s$ Weyl tensor which uniquely
defines the full $B$-propagator via a simple integral map. We have
applied this machinery to explicitly find all the boundary-to-bulk
$B$-propagators.

The dynamics of higher spin fields is described in terms of differential equations and algebraic constraints that set the fields on-shell. The algebraic constraints are quite complicated to work with in perturbation theory. Fortunately, in lower dimensions most complicated part of the algebraic constraints can be easily resolved by introducing twistor-like variables instead of vector-like. It is because of these simplifications the computations of three-point functions in $4d$ higher-spin theory are quite simple, \cite{Giombi:2010vg}. Unfortunately, no analog of this twistor resolution is known in arbitrary dimension.

The paper is organized as follows. In Section \ref{SecFromScalartoVasiliev} we show a simple route from the free conformal scalar on the boundary to the higher-spin theory in the bulk, the goal being to show that it is the higher-spin master connection $W$ that couples naturally to the currents built of scalar fields. In Section \ref{SecAmbient} we briefly review the ambient approach, more specific for our problem discussion is presented in Section \ref{SecGeometry}. The linearized Vasiliev higher-spin equations \cite{Vasiliev:2003ev} are reviewed in Section \ref{SecVasilievEquations}. Boundary-to-bulk propagators for master connections and field-strengths are derived in Sections \ref{SecWpropagator} and \ref{SecBpropagator}, where two-point functions are also discussed. Conclusions are collected in Section \ref{SecConclusions}.

\section{From boundary to higher-spin fields in the bulk}\label{SecFromScalartoVasiliev}
Here we would like to present a short path to the gauge fields
introduced by Vasiliev in \cite{Vasiliev:2001wa} and then used in
\cite{Vasiliev:2003ev} to construct a classical theory of
interacting higher-spin fields in $AdS_{d+1}$. One starts with an $u(N)$, $so(N)$, ... multiplet $\phi^I(x)$ of free scalar fields in flat space of dimension $d$.
As is well known, having two $\phi^I(x)$ and $s$ derivatives one
can construct a conserved current that is a traceless tensor of
any rank $s=1,2,3,...$\footnote{This is true for $u(N)$. In the case
of $so(N)$ there exist currents of even ranks only. Indices $\mu,\nu,...$ run over $d$ values. A group of $s$ (anti)symmetric or to be (anti)symmetrized indices $\mu_1...\mu_s$ is denoted by (\,$\mu[s]$\,) $\mu(s)$}, the trace over the vector indices being implicit,
\bee\label{SpinSCurrent}
&j_{\mu(s)}=\phi(x) (\overleftarrow{\pl_\mu}-\overrightarrow{\pl_\mu})^s\phi(x)-\mbox{traces}, \qquad\qquad\qquad& \pl^\nu j_{\nu\mu(s-1)}=0\,.
\eee
For $s=2$ one recognizes usual stress-energy tensor. The conformal dimension of a spin-$s$ current is $2\Delta+s$, where $\Delta=(d-2)/2$ is the dimension of a free scalar. As well as
$\phi^I(x)$ themselves, these currents form a representation of
the conformal algebra $so(d,2)$, which means that the charges are
labeled by certain modules of $so(d,2)$. Indeed, the conservation
condition
\bee
&\pl^\mu j^s_{\mu}=0,\qquad\qquad\qquad\qquad\quad&j^s_\mu=j_{\mu}{}^{\nu(s-1)} K_{\nu(s-1)}(x)\,,
\eee
for yet unknown traceless $K_{\nu(s-1)}(x)$ implies that
\bee \mbox{traceless part of }[\pl^{\nu}K^{\nu(s-1)}]=0\,,
\eee
the latter equation being (i) conformally covariant, (ii) overdetermined, as such admitting a finite number of solutions. Its solutions are called the conformal Killing tensors (CKT), which generalize conformal Killing vectors. Any given CKT allows one to define a conserved charge in a standard way as an integral of a $(d-1)$-form
\bee
Q=\int \Omega^s, \qquad\qquad\qquad \Omega^s=j^s_\mu\,\epsilon^{\mu}{}_{\sigma[d-1]}\, dx^{\sigma}\wedge...\wedge dx^\sigma\,.
\eee
(i)+(ii) implies that CKT are just $so(d,2)$-tensors, although
this is not easy to see, \cite{Eastwood:2002su}. Namely, various components in the Taylor
expansion of $K_{\nu(s-1)}(x)$ can be organized into an
irreducible $so(d,2)$-tensor\footnote{Indices $A,...$ run over
$d+2$ values $\la,+,-$, where $\la,\lb,...$ are fiber $so(d-1,1)$ indices. $so(d,2)$-irreducibility
means that (1) the tensor indices have a symmetry of some Young
diagram, (2) the tensor is traceless. } $K^{A(s-1),B(s-1)}$ that
has a symmetry of a two-row rectangular Young diagram of
length-$(s-1)$
\bee\label{TwoRowRectangular}
\parbox{2.3cm}{\bep(60,20)\put(0,0){\RectBRow{6}{6}{$s-1$}{$s-1$}}\put(0,0){\YoungAA}\put(50,0){\YoungAA}\eep}\,.
\eee
The map from the explicitly conformal $K^{A(s-1),B(s-1)}$ to the hiddenly conformal CKT $K_{\nu(s-1)}(x)$ reads \cite{Eastwood:2002su}
\bee
&K_{\nu(s-1)}(x)=M^{AB}_\nu...\,M^{AB}_\nu K_{A(s-1),B(s-1)},& \\
&M^{AB}_\nu = X^A \pl_\nu X^B-X^B \pl_\nu X^A,\qquad\qquad\qquad& X^A=\{1,x^\la, -x^\la x_\la/2 \}\,,
\eee
where $X^A(x)$ is a Poincare slice of the zero cone $X^A X_A=0$
(see Section \ref{SecAmbient} below). The generators $M^{AB}$ decompose
into dilatation $D=M^{+-}$, translations $P^\la=M^{+\la}$, Lorentz
rotations $L^{\la\lb}=M^{\la\lb}$ and conformal boosts $K^\la=M^{-\la}$. In
particular it is evident that a Killing tensor gets decomposed
into a product of Killing vectors. The Killing vectors associated
to $P^\la$, $D$, $L^{\la\lb}$ and $K^\la$ read
\bee
P^\la_\nu&=\delta^\la_\nu\,, \qquad\qquad& D_\nu= x_\nu\,,\\
L^{\la\lb}_\nu &= \delta^\la_\nu x^\lb-\delta^\lb_\nu x^\la\,, \qquad\qquad &
K^\la_\nu = 2x_\nu x^\la-\delta^\la_\nu x^\lm x_\lm\,.
\eee

Therefore, the on-shell closed form $\Omega$, defining the whole conformal multiplet of the charges associated with $j_{\mu(s)}$, is naturally a carrier of the label of a Killing tensor,
\bee
\Omega=\Omega^{A(s-1),B(s-1)}K_{A(s-1),B(s-1)}\,.
\eee

The last but one step is to couple currents $j_{\mu(s)}$ to some external fields, $\phi^{\mu(s)}$, the Fradkin-Tseytlin fields \cite{Fradkin:1985am}, which are gauge fields as the currents are conserved,
\bee\label{FradkinTseytlin}
&\Delta S =\int \phi^{\mu(s)}\, j_{\mu(s)}, \qquad\qquad&\delta\phi^{\mu(s)}=\pl^\mu\xi^{\mu(s-1)}-\mbox{traces}\,.
\eee
Such a coupling, however,  involves only one component of the
whole $so(d,2)$-multiplet, the one associated with the
$P^a...P^a$-part of the conformal Killing tensor
$K_{A(s-1),B(s-1)}$.

It is now natural to introduce a gauge field that draws in all the
components of the multiplet, such conformal fields were considered in \cite{Vasiliev:2009ck},
\bee\fl
&\Delta S =\int \Omega_{A(s-1),B(s-1)}\wedge W^{A(s-1),B(s-1)},\qquad \delta W^{A(s-1),B(s-1)}=D  \xi^{A(s-1),B(s-1)}\,.
\eee
It has to be a one-form since $\Omega$ is a $(d-1)$-form and the gauge parameter $\xi$ is a zero-form
\bee
\qquad\qquad W^{A(s-1),B(s-1)}\equiv W^{A(s-1),B(s-1)}_\mu\, dx^\mu \label{VasilievConn}
\eee
that take values in the same irreducible $so(d,2)$-module as $\Omega$ does. $D$ is a flat covariant derivative of $so(d,2)$. Then, $\phi^{\mu(s)}$ is associated with one particular component of $W$,
\bee\label{phifield}
\phi_{\mu}{}^{\la(s-1)}=\mbox{totally symmetric and traceless part of } W_\mu{}^{\la(s-1),++...+}
\eee

Let us note that the current associated with $K^{a(s-1),++...+}$ plays a distinguished role, of course, as it is the highest-weight current in the multiplet and all other currents can be thought of as its descendants. However, the importance of having some symmetry manifest rather than playing with a small part of it should not be underestimated.

For the case $s=2$, i.e. for the energy-momentum tensor
$j_{\mu\mu}$, this tells us that $W=W^{A,B}_\mu\, dx^\mu$ is a
usual Yang-Mills connection of the conformal algebra and it
couples naturally to all currents that can be built out of
$j_{\mu\mu}$ with Killing vectors associated to the generators $D,
P^a, L^{ab}, K^a$ of the conformal group.

The last step is to interpret $W$ in the spirit of AdS/CFT
correspondence as a boundary value of a bulk field, which might
be called the higher-spin gauge connection, \cite{Vasiliev:2001wa}. Taking
into account that there are currents of all spins (at least of
even ranks) in the theory of free scalars, the dual theory is
expected to be a gauge theory of all connections of type
(\ref{VasilievConn}) and it does exist \cite{Vasiliev:2003ev}.

\section{Bulk and boundary in ambient space}\label{SecAmbient}
A relation between theories in anti-de Sitter space and their
conformal partners should become more transparent and less
technically involved if both types of theories are put into the
same space where the conformal symmetries/anti-de Sitter global
symmetries act geometrically. This is the ambient space, a flat
pseudo-euclidian space $\mathbb{R}^{d,2}$, see
\cite{Dirac:1935zz, Dirac:1936fq, Fronsdal:1978vb, Eastwood:1991, Metsaev:1995re,
Metsaev:1997nj, Preitschopf:1998ei, Bars:1998ph, Barnich:2006pc,
Boulanger:2008up, Bekaert:2009fg, Costa:2011mg,
SimmonsDuffin:2012uy, Bekaert:2012vt} for original works and developments. Below we
introduce the notation we need rather than reviewing the ambient
approach.

\paragraph{\bf Bulk.} The anti-de Sitter space $AdS_{d+1}$ is understood by definition as the hyperboloid \bee X^A X_A\equiv X\cdot X=-R^2\eee and we put $R=1$ for simplicity. It is obvious that any $SO(d,2)$ rotation $\Lambda\fud{A}{B}$ preserves the hyperboloid, so the representation of $SO(d,2)$ is simply $\rho(\Lambda)X=\Lambda\fud{A}{B} X^B$.

A symmetric tensor field of the Lorentz algebra $so(d,1)$ is
defined on AdS as a tensor field $\phi^{A(s)}(X)$ satisfying, $X_B
\phi^{BA(s-2)}=0$, with the latter condition properly reducing the
number of independent components. So the local Lorentz algebra at
$X$ is defined as the stability algebra of $X$. One can assume
that the fields live on $X^2=-1$ or extend them in radial
direction by imposing certain conditions, e.g. homogeneity,
see \cite{Bekaert:2012vt} for a recent progress.

The Lorentz covariant derivative is defined as
\bee
D_M \phi^{A(s)}= G\fud{N}{M} G\fud{A}{B}...G\fud{A}{B}\, \pl_N \phi^{B(s)},\qquad G_{AB}=\eta^{AB}+X^AX^B\,,
\eee
where $G$ is a projector onto $X$-transverse subspace in the tangent space and it brings the indices back to the Lorentz subspace. $G$ also serves as the ambient realization of the Lorentz metric and as a vielbein. Despite the unusual from of $D_M$ one can verify that it amounts to the usual Lorentz covariant derivative in any local coordinates. A distinguished parameterization of AdS is given by Poincare coordinates
\bee\label{PoincareAdS}
X^A=\frac1{x^0} (1,x^\la, -x^\lm x_\lm/2-(x^0)^2/2)\,.
\eee

\paragraph{\bf Boundary.} The conformal boundary of AdS is understood as a projective light-cone,
\bee
Z^A Z_A=0\,,\qquad \qquad Z^A\sim \lambda Z^A, \qquad \lambda\neq0\,.
\eee
That $Z$ is null allows one to impose an additional factorization condition. In order to work with the equivalence relation effectively one may choose a gauge, which can be imposed with the help of an auxiliary vector $V$. A distinguished gauge $V\cdot Z=1$ with $V^A=\delta^A_-$ leads to the Poincare slice of the cone
\bee\label{PoincareCone}
Z^A=(1,z^\la, -z^2/2)\,.
\eee
Then $SO(d,2)$ becomes acting as $\rho(\Lambda)Z=\Lambda\fud{A}{B} Z^B/(V\Lambda Z)$\,.

A symmetric tensor field $T^{A(s)}(Z)$ is a conformal quasi-primary if: (i) it is homogeneous of some degree $\delta$, $T(\lambda Z)=\lambda^{-\delta} T(Z)$, which is the conformal weight; (ii) it is irreducible, i.e. it is traceless; (iii) it is transverse to the cone, i.e. $Z_B T^{BA(s-1)}=0$; (iv) it is defined modulo gauge transformations $\delta T^{A(s)}=Z^A \xi^{A(s-1)}$. (ii)+(iii)+(iv) reduce the number of independent components to that of an irreducible $so(d-1,1)$ tensor. Again one might wish to extend the fields off the cone by imposing further restrictions on $Z$-dependence. (ii)+(iii)+(iv) can be encoded by contracting the indices with a polarization vector $\eta^A$, such that $\eta\cdot \eta=0$, $Z\cdot \eta=0$, $\eta\sim\eta+Z$. Various conformal structures, e.g. the ones appearing in the correlators can be effectively written in the ambient space, \cite{Costa:2011mg}.

Let us note that up to Fourier transform and the scale condition,
the tensor fields on AdS are just massive fields in
$\mathbb{R}^{d,2}$, for which AdS is a mass-shell, while fields on
the boundary are just massless fields in $\mathbb{R}^{d,2}$, for
which the cone is a mass-shell and the extra equivalence relations
are just usual gauge symmetries of massless fields.

\section{Geometry of boundary-to-bulk problem in ambient space}\label{SecGeometry}
It is useful to list, see Table \ref{TblGeometric} and picture
\ref{PicMain} below, all the variables that are relevant for the
boundary-to-bulk problem where a source field on the boundary is a
totally-symmetric traceless tensor, which as mentioned above can
be encoded with the help of a null polarization vector $\eta$, so
we will consider polynomials in $\eta$ instead.
\begin{figure}[h!]
  \caption{Geometry of boundary-to-bulk problem in ambient space}
  \centering
 \includegraphics[height=8cm, width=13cm]{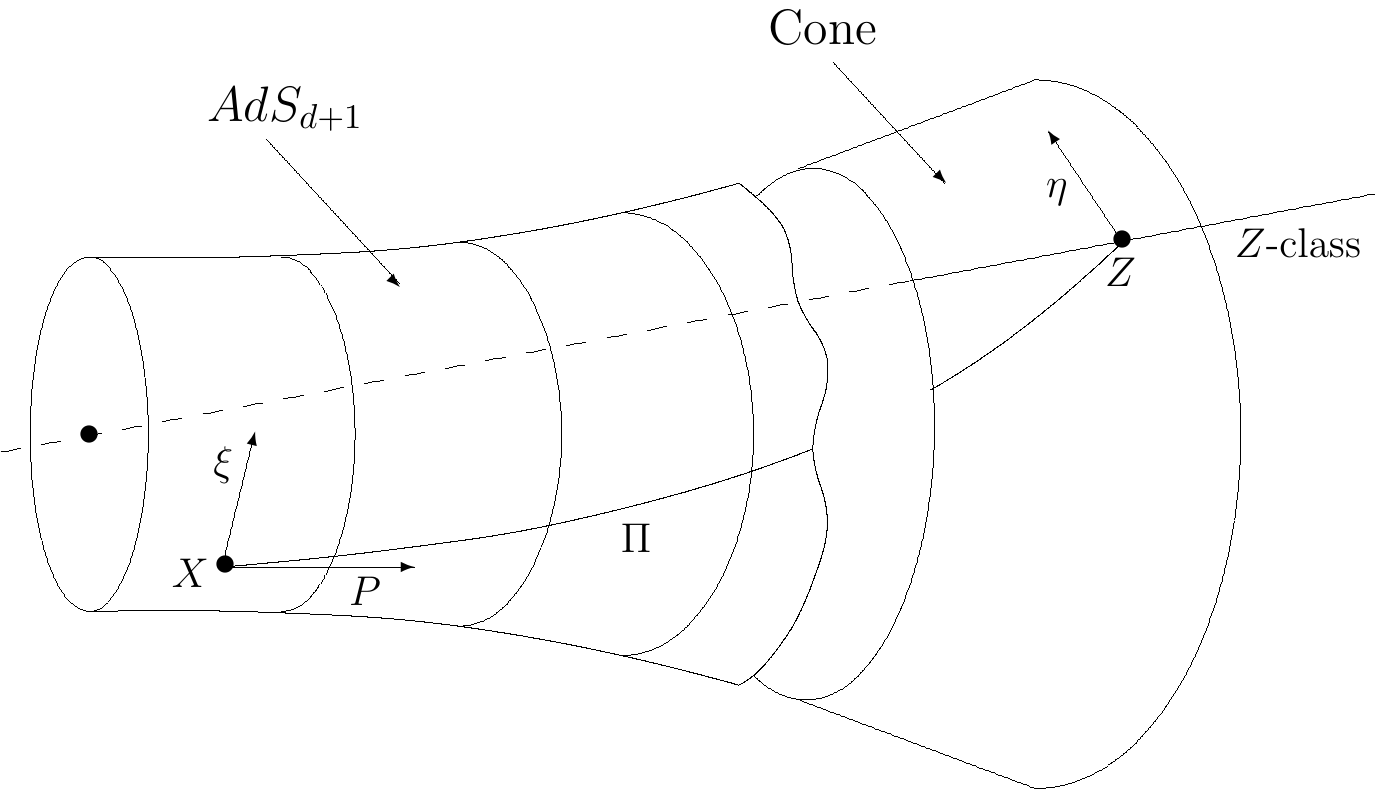}
 \label{PicMain}
\end{figure}

\begin{table}[h]
\caption{Relevant geometric quantities for the boundary-to-bulk problem}
\begin{tabular}{|c|w{9.6cm}|}\hline
quantity&\centering description \tabularnewline\hline\hline $X^A$
& a point in the bulk, i.e. on the hyperboloid, $X^2=-1$\tabularnewline\hline $Z^A$\rule{0pt}{12pt} & a point on the boundary, i.e.
on the cone, $Z^2=0$\tabularnewline\hline \rule{0pt}{16pt}
$(X Z)$ & the 'geodesic distance' between $X$ and
$Z$\tabularnewline\hline $\eta^A$ & polarization vector on the
boundary at point $Z$, i.e. $\eta\cdot \eta=0$, $\eta\cdot Z=0$,
$\eta\sim\eta+Z$\tabularnewline\hline
\rule{0pt}{16pt}$\refp{X}{G}_{AA}=\eta_{AA}+X_A X_A$ & induced Lorentz
metric on AdS, $G_{AB}X^B=0$. It also plays the role of the frame
field $E^A_\wiM=G\fud{A}{\wiM}$ in the ambient space.
\tabularnewline\hline $\refp{X}{P}_A=X_A+Z_A/(X Z)$ & a vector that is
tangent to the 'geodesic' connecting $X$ with a boundary point $Z$.
It is a 'wave-vector' of a plane-wave towards $Z$. It respects the equivalence class of $Z$
\tabularnewline\hline $\Pi(X,Z)\fud{A}{B}=\delta\fud{A}{B}-\frac{\displaystyle Z^A
X_B}{\displaystyle(X Z)}$ & a parallel transport tensor, that propagates tensor
indices from the boundary point $Z$ to the bulk point $X$. It respects the equivalence classes of $Z$ and $\eta$
\tabularnewline\hline
\rule{0pt}{16pt}$\refp{X}{\xi}^A=\Pi(X,Z)\fud{A}{B}\, \eta^B$ & the
polarization vector $\eta$ that is parallel transported to the
bulk point $X$. In addition to being tangent $\xi \cdot X=0$ and
null $\xi \cdot \xi=0$ it is also orthogonal to the 'wave-vector'
$P$, $\xi \cdot P=0$\tabularnewline\hline
\end{tabular}
\label{TblGeometric}
\end{table}

To do computations it is important to have explicitly all the
derivatives of the quantities given in the Table \ref{TblGeometric}, fortunately
these are closed on themselves and are given below  (we omit the reference point superscript, e.g. just $P_A$ instead of $\refp{X}{P}_A$)
\bee\fl\qquad\eqalign{\label{DerivativesAll}
D_A(XZ)=(XZ) P_A\,, \qquad\qquad& D_A P_B=G_{AB}-P_A P_B\,, \\
D_A\xi_B=-\xi_A P_B\,,  &D_M G_{AB}=0\,,\qquad\qquad\qquad D_M X^A=0\,.}
\eee

\paragraph{\bf Simplest boundary-to-bulk propagators.} To get the feeling that the ambient approach makes things simpler let us consider 1) propagators for scalars and 2) totally-symmetric fields.

\paragraph{\bf1. Scalar b-to-b.} From the pioneering AdS/CFT works \cite{Gubser:1998bc, Witten:1998qj, Freedman:1998tz} the boundary-to-bulk propagator for a weight $\delta$ scalar is up to normalization simply
\bee\fl\qquad\eqalign{\label{ScalarProp}
K_\delta(X|Z)=\frac1{( X\cdot Z)^\delta}\,,\qquad\qquad\qquad (D^2-\delta(\delta-d))K_\delta(X|Z)=0\,,\\
-\frac12(X Z)^{-1}=\frac{x^0}{(x^0)^2+(x-z)^2}\quad(\mbox{in Poincare coordinates})}
\eee
this is so because $\pl_C \pl^C K_\delta\sim Z^2=0$.
\paragraph{\bf2. Spin-s b-to-b.} Starting form spin-one the boundary-to-bulk propagators become more and more complicated in intrinsic coordinates, \cite{D'Hoker:1998gd, Liu:1998ty, D'Hoker:1999jc}. A totally-symmetric spin-$s$ field, whose boundary value is the Fradkin-Tseytlin field (\ref{FradkinTseytlin}) that couples to the current $j_{\mu(s)}$, can be described \cite{Fronsdal:1978rb} by the Fronsdal field $\phi^{A(s)}(X)$ that obeys
\bee\eqalign{\label{FronsdalEqs}
&\fl(D^2-m^2) \phi^{A(s)}-sD^AD_M\phi^{A(s-1)M}+\frac{s(s-1)}2\left(D^AD^A-2G^{AA}\right)\phi\fud{A(s-2)C}{C}=0\,,\\
&\fl m^2=E(E-d)-s \,\qquad\qquad\qquad\qquad\qquad G_{BB}G_{BB}\phi^{A(s-4)B(4)}\equiv0\,,\\
&\fl \delta \phi^{A(s)}(X)=D^A \xi^{A(s-1)}\,,\qquad\qquad\qquad\qquad\quad G_{BB}\xi^{A(s-3)BB}\equiv0\,,}
\eee
where $E=2\Delta+s$ is the lowest energy of the field, \cite{Metsaev:1997nj}. It coincides with the dimension of the spin-$s$ current $j_{\mu(s)}$, (\ref{SpinSCurrent}). Such a gauge field with somewhat strange double-trace constraints comes naturally as a part of the
higher-spin connection, \cite{Vasiliev:2001wa}. A propagator for a spin-$s$ field was
proposed in \cite{Mikhailov:2002bp}, when slightly refined it reads
\bee\eqalign{\label{MikhailovProp}
&K_\delta(X|Z,\eta)^{A(s)}=\frac1{(X Z)^\delta}\xi^{A(s)}\,, \qquad\qquad G_{BB} K_\delta^{A(s-2)BB}=0,\\
&D_M K_\delta^{A(s-1)M}=0\,,\qquad\qquad\qquad(D^2-\delta(\delta-d)+s)K_\delta^{A(s)}=0\,,}
\eee
where the above conditions are satisfied for any $\delta$, but only for $\delta=2\Delta+s$ $K_\delta$ is a propagator for the Fronsdal field we will need. (\ref{MikhailovProp}) is the lowest part of the higher-spin master field $W$ that is a part of the Vasiliev formulation, \cite{Vasiliev:2003ev}.

\paragraph{\bf Boundary limit prescription.} Within the context of AdS/CFT it is important to know the limit of all geometric quantities introduced above.
There are two natural prescriptions. First is to recover $X^2$
factors, which were dropped as $X^2=-1$, and then try to take
$X^2\rightarrow0$.
This is more complicated, however. Second is to use the
Poincare coordinates experience. Naively, that $X$ grows near the boundary as
${\varphi}^{-1}$, where $\varphi\sim z$ is a defining function of
the conformal boundary, suggests to the leading order
\bee\fl\quad
P^A\rightarrow X^A\,,\qquad\qquad G^{AA}\rightarrow X^A X^A\,,\qquad\qquad \Pi(X,Z)\rightarrow \Pi(\overline{X},Z)\,,
\eee
where $\Pi(X,Z)$ is formally unchanged, but $X$ gets replaced by the Poincare slice $\overline{X}$, (\ref{PoincareCone}), of the cone, $\overline{X}=\lim z X$.

As is well known, \cite{Balasubramanian:1998de}, given a boundary-to-bulk propagator
$K_\delta$, which tends to $(x^0)^{d-\delta}\delta(x-z)$, the
coefficient of the second asymptotic $(x^0)^\delta$ is directly
proportional to the two-point function. From
\bee
K_\delta=\frac1{(X Z)^\delta}=\frac{(x^0)^\delta}{(\overline{X}\cdot Z)^\delta} \sum_k \frac{\Gamma[\delta+k]}{\Gamma[\delta]}\frac{(-)^k (x^0)^{2k}}{(-2 \overline{X}\cdot Z)^{k}}\qquad
\eee
one observes that this is the coefficient of $1/(\overline{X}\cdot
Z)^\delta$, i.e. the two point functions up to some numerical but
still important factor are obtained just by changing the meaning
of $X$ to be that of a point on the cone.

With the above prescription one sees that (\ref{ScalarProp})
immediately gives $\langle \phi(X)\phi(Z)\rangle=(X Z)^{-\delta}$.
Introducing an auxiliary polarization $\zeta^A$ at point $X$ to
contract the indices of the boundary-to-bulk propagator (\ref{MikhailovProp}) for the Fronsdal field one finds
\bee\label{TwoPointCurrent}
\langle j_{s_1}(X) j_{s_2}(Z)\rangle=\delta_{s_1,s_2}\frac{1}{(X Z)^{2\Delta}} \left(\frac{\zeta^A\, \Pi(X,Z)_A{}^B\, \eta_B}{(X Z)}\right)^s
\eee
which is a correct expression for the two-point function of conserved currents, \cite{Costa:2011mg}.

\paragraph{\bf Twisted-adjoint action of the conformal group.} Below it will be also important
to have a somewhat unusual action of $so(d,2)$ on $\xi_A$ and
$P_A$. For the reasons that become clear in Section
\ref{SecBpropagator}, we call this the twisted-adjoint action.
Let the polarization $\refp{V}{\xi}$ and 'wave-vector'
$\refp{V}{P}$ be given at some bulk point $V^A$. Then an $AdS$
rotation $\Lambda\fud{A}{B}$ is performed that takes $V$ to $X$. The boundary
point $Z$ is kept fixed by hand, i.e. we would like to have a
transformation that takes $\refp{V}{P}(V,Z)$ to $\refp{X}{P}(X,Z)$
({\it idem.} for $\xi$) without acting on $Z$, which means that
the hyperboloid is rotated while the cone does not. The
corresponding transformations read
\bee\eqalign{\label{TwistedAdjointRot}
X^A&=\Lambda\fud{A}{B}V^B\,, \\
\refp{X}{P_A}&=\frac1{\sigma} (\refp{V}{P_A}-V_A)+X_A\,, \qquad\qquad\qquad \sigma= (\refp{V}{P}-V)\cdot X\,,\\
\refp{X}{\xi}_A&=\refp{V}{\xi}_{A}-\frac1{\sigma}
(\refp{V}{P}_A-V_A) (\refp{V}{\xi}\cdot X)\,.}
\eee

\section{Higher-spin fields}\label{SecVasilievEquations} In this section we present the equations that describe
free higher-spin fields in anti-de Sitter space of any dimension
$d+1$ in terms of certain {\it master} fields. The basic material
is of course well-known, e.g. see \cite{Vasiliev:2001wa,
Eastwood:2002su, Vasiliev:2003ev, Bekaert:2005vh,
Iazeolla:2008ix}, but the exposition is somewhat new. Firstly, the
master fields that take values in the higher-spin algebra are
introduced. Secondly, the background geometry, i.e. the anti-de
Sitter space, is given in a form analogous to higher-spins
themselves. Thirdly, the equations are presented and few
properties thereof are discussed. At the end an effective
oscillator realization is reviewed.

{\bf1. Algebra.} We start with the generators $T_{AB}$ of the anti-de Sitter or conformal algebra $\mathfrak{h}=so(d,2)$
\bee\label{SOCommuRelations}
[T_{AB},T_{CD}]_{\star}&=T_{AD}\eta_{BC}-T_{BD}\eta_{AC}-
T_{AC}\eta_{BD}+T_{BC}\eta_{AD}\,,
\eee
where $\star$ is the product in the universal enveloping algebra $U(\mathfrak{h})$ of $\mathfrak{h}=so(d,2)$. What we would like to review is that there exists an algebra $\mathfrak{g}$, called higher spin algebra, whose connection $W(T|X)$ gets decomposed under $\mathfrak{h}$ in terms of connections (\ref{VasilievConn}) whose boundary values couple to all currents build of free scalar fields, i.e.
\bee\fl
W(T|X)=\sum_s W^{A(s-1),B(s-1)}\, T_{AB}\star...\star
T_{AB}\equiv\sum_s W^{A(s-1),B(s-1)}\,
T_{AB(s-1)}\,.\label{VasilievMasterConnection}
\eee
$U(\mathfrak{h})$ is a good starting point as it is
a quite large extension of $\mathfrak{h}$, which should have enough room. $U(\mathfrak{h})$ is an $\mathfrak{h}$ module itself, whose decomposition in terms of irreducible $\mathfrak{h}$-modules can be worked out using the Poincare--Birkhoff--Witt theorem, the first levels being given by
\bee\fl\label{SOspectrum}
\left. U(\mathfrak{h})\right|_{\mathfrak{h}}
&\cong \underbrace{\bullet}_{0}\oplus\underbrace{\left(\;\parbox{10pt}
{\YoungAA}\;\right)}_1\oplus
\underbrace{\left(\parbox{20pt}{\YoungBB}\oplus\parbox{10pt}
{\YoungAAAA}\oplus\parbox{20pt}{\YoungB}\oplus\bullet\right)}_2
\oplus
\underbrace{\left(\;\parbox{30pt}{\YoungCC}\oplus\ldots\;\right)}_3\oplus
\ldots
\eee
where the first singlet $\bullet$ is just the unit of
$U(\mathfrak{h})$, $\parbox{10pt}{\YoungAA}$ represents $T^{AB}$. At the level two, the singlet is the quadratic
Casimir operator $C_2=-\frac12T_{AB}\star T^{AB}\,$ and two more elements
\bee\label{Idealtofactor}
&\parbox{10pt}{\YoungAAAA}=T_{[AB}\star T_{CD]}, \qquad\qquad&
\parbox{20pt}{\YoungB}=T\fdu{A}{C}\star T\fdu{AC}{}
-\frac{2}{(d+1)}\eta_{AA}C_2
\eee
are the first ones that do not fit into the pattern of
(\ref{VasilievConn}), (\ref{VasilievMasterConnection}) as they do
not have the symmetry of a rectangular two-row Young diagram,
(\ref{TwoRowRectangular}). It is necessary to quotient them out,
defining a two-sided ideal $I$
\bee
&I\cong
U(\mathfrak{h})\star\left(\;\parbox{10pt}{\YoungAAAA}\oplus\parbox{20pt}
{\YoungB}\;\right)\star U(\mathfrak{h})\quad\,.\eee Despite not
being immediately obvious, the procedure is consistent and there
is nothing else to care about (see \cite{Eastwood:2002su,
Vasiliev:2003ev, Bekaert:2005vh, Iazeolla:2008ix,
Boulanger:2011se} for an extended elaboration), the higher-spin
algebra defined by $\mathfrak{g}=U(\mathfrak{h})/I$ has the
desired decomposition in terms of $\mathfrak{h}$-modules \footnote{ All the other unwanted diagrams in
the spectrum turns out to be removed by $I$. Actually $I$ is the
annulator of a free conformal scalar, \cite{Eastwood:2002su, Konshtein:1988yg,
Iazeolla:2008ix}.}.
\bee\label{HSAlgebraDecomposition}
\mathfrak{g}\cong\bullet\oplus\parbox{10pt}
{\YoungAA}\oplus\parbox{20pt}{\YoungBB}\oplus\parbox{30pt}{\YoungCC}
\oplus\ldots
\eee
In addition to the master one-form connection $W(T|X)$ one should introduce the field-strengths that are packed into the master zero-form field $B(T|X)$.

From the Fronsdal $\phi^{A(s)}$ field (\ref{FronsdalEqs}) vantage
point, the $W(T|X)$-field will encode non-gauge invariant
derivatives of $\phi^{A(s)}$, while $B(T|X)$ will encode gauge
invariant ones. Introduction of auxiliary fields to encode
the derivatives of the $\phi^{A(s)}$ field is the matter of
convenience as the higher-spin interactions contain higher-order
derivatives. One then may split a problem of interactions into
(i) writing constraints that encode all derivatives of
$\phi^{A(s)}$ in terms of master fields; (ii) looking for purely
algebraic couplings of master fields that preserve the
constraints, \cite{Vasiliev:2003ev}.

{\bf2. Geometry. } The anti-de Sitter space can be defined via a
flat connection $\Omega=\frac12 \Omega^{A,B}\,T_{AB}$,
$\Omega^{A,B}\equiv \Omega^{A,B}_\mu dx^\mu$, which is the vacuum
value of $W(T|X)$,
\bee
d\Omega+\Omega\star\Omega=0\label{OmegaFlat}\,.
\eee
In order to define the notion of a Lorentz tensor at any point of
the anti-de Sitter space one should introduce \cite{Stelle:1979aj,
Vasiliev:2001wa} an external field $V^A(X)$, $V\cdot V=-1$, called
compensator, that defines the splitting of the local $so(d,2)$
into the Lorentz subalgebra $so(d,1)$, which is a stability subalgebra of $V$, and translations, c.f. Section \ref{SecAmbient}. Roughly speaking this
amounts to splitting $\Omega^{A,B}$ into vielbein $h^a_\mu$ and
spin-connection $\varpi^{a,b}_\mu$, which as a consequence of
(\ref{OmegaFlat}) will satisfy\footnote{Indices $a,b,..$ are fiber indices of the AdS Lorentz algebra $so(d,1)$. $\Lambda$ is the cosmological constant.}
\be\label{OmegaInLorentz}
d\varpi^{a,b}+\varpi^{a,}{}_{c}\wedge\varpi^{c,b}=-\Lambda h^{a}\wedge
h^{b}\,,\qquad dh^{a}+\varpi^{a,}{}_{c}\wedge h^{c}=0\,.
\ee
This can be done in a fully $so(d,2)$-covariant way as in
\cite{Stelle:1979aj, Vasiliev:2001wa}, the expressions for the
vielbein $E^A$ and Lorentz-covariant derivative $\DL$ read
\bee
&E^A=dV^A+\Omega\fud{A}{B}V^B\,, & E^A V_A=0\,,\\
&\DL=d+\frac12 T_{AB}\,(\Omega^{A,B}+\Sigma^{A,B})\,, \qquad& \Sigma^{A,B}=E^AV^B-E^BV^A\,.
\eee
The Lorentz-covariant derivative $\DL$ is determined by $\DL V^A=0$,
$\DL E^A=0$. It is worth mentioning that $E^A_M dX^M$ must have the
maximal rank, i.e. $(d+1)$, which is a standard requirement for the
vielbein, otherwise one cannot interpret the theory given below in
terms of Lorentz tensors.

A standard choice for the compensator is $V^A=\mbox{const}=\delta^A_{d+1}$. Let us note that within the ambient approach it is natural to choose $V$ be just an ambient coordinate $X$, then
the vielbein is $E\fud{A}{M}\, dX^M=G\fud{A}{M}\, dX^M=\pl_M X^A\, dX^M$. The local Lorentz generators are the $V$-orthogonal components of
$T^{AB}$. The translation generators are simply $P^A=T^{AB}V_B$.

{\bf3. Equations. } The equations that describe free higher-spin fields read
\numparts
\bee\label{EqLorentzA}
&\DL W-\frac12\Sigma_{A,B}\,[T^{AB}, W]_\star=\left.E^A\wedge E^B \ddd{T^{AB}} B\right|_{P^A=0}\,,\\
&\DL B-\frac12\Sigma_{A,B}\,\{T^{AB}, B\}_\star=0\label{EqLorentzB}\,.
\eee
\endnumparts
Let us mention briefly several important properties of these equations.

{\bf a.} The equations are consistent and complete (integrable) in a
sense that applying $d$ and using the equations again gives zero
and does not produce any new constraints on the fields. Equations
have the unfolded form \cite{Vasiliev:1988xc, Vasiliev:1988sa}, i.e. are of the first order, written by
making use of the exterior products of differential forms and de Rham differential $d$, which now is hidden inside the Lorentz covariant derivative $\DL$. The unfolded
equations enjoy a number of nice properties \cite{Vasiliev:1999ba, Vasiliev:2007yc, Gelfond:2008td}. In particular all one-forms,
i.e. $W$ in our case, take values in some Lie algebra, which
follows from the integrability requirement. Then, all the
structures appearing in the unfolded equations have an
interpretation in terms of this Lie algebra.

{\bf b.} The full
nonlinear equations for higher-spin fields \cite{Vasiliev:2003ev}
are given in the unfolded form and are certain nonlinear
deformations of (\ref{EqLorentzA})-(\ref{EqLorentzB}).

{\bf c.} If
it were not for the {\it r.h.s.}, (\ref{EqLorentzA}) would be a
covariant constancy condition in the adjoint representation of the
higher-spin algebra. Its {\it l.h.s.} is simply
        \bee
        &\DO\,W =d W+[\Omega, W]_\star=...
        \eee
      which from the point of view of nonlinear theory is to be understood as a linearization of $dW+W\star W$ over the $\Omega$ background. At the linearized level the higher-spin algebra is just a highly reducible $so(d,2)$-module, see (\ref{HSAlgebraDecomposition}), as $\Omega$ does not have any components beyond $so(d,2)$.

{\bf d.} Equations (\ref{EqLorentzA})-(\ref{EqLorentzB}) decompose under $so(d,2)$ into independent set of equations, one set for each spin $s=0,1,2,...$. A scalar field has all its derivatives in $B$ field, while any $0<s$-field has its derivatives split between $W$ and $B$.

{\bf e.} Importantly, the consistency of the equations is not spoiled
by the {\it r.h.s.} of (\ref{EqLorentzA}), which has an
interpretation as a Chevalley-Eilenberg cocycle of $so(d,2)$.

{\bf f.} If it were not for the {\it r.h.s.} of (\ref{EqLorentzA}) the equation $dW+W\star W=0$ would have pure
gauge solutions only, $g^{-1}\star dg$, describing no propagating degrees of freedom in the bulk. It is the gluing term
that makes $W$ propagating and it is the difficulty to deform the gluing term that makes the higher-spin problem so complicated. That $P^A=0$ on the
{\it r.h.s.} of (\ref{EqLorentzA}) tells us that not all
components of $W$ are sourced by $B$, but only those that are
transverse to the compensator, these are called Weyl tensors. Schematically (\ref{EqLorentzA}) reads
\bee\fl\label{WeylTensorDef}
dW^{A(s-1),B(s-1)}+...=E_M \wedge E_N C^{A(s-1)M,B(s-1)N},\qquad\qquad V_M C^{A(s-1)M,B(s)}\equiv0\,.
\eee

{\bf g.}
(\ref{EqLorentzB}) is a covariant constancy equation, but given
with respected to the twisted-adjoint action of $so(d,2)$. Given any
automorphism $\pi$ one can define a twisted-adjoint action
$T(B)=T\star B-B\star \pi(T)$, which is still a representation of
the algebra. In the higher-spin case, $\pi$ reflects the
local translation generators $P^A=T^{AC}V_C$ while not affecting
the Lorentz ones, explicitly,
      \bee\fl
        \pi(T^{AB})=T^{AB}+2P^{A} V^B-2P^{B} V^A,\qquad\qquad \pi^2=id,\qquad \pi(P^A)=-P^A\,.
      \eee
      One may rewrite (\ref{EqLorentzB}) in a more $\Omega$-covariant form, emphasizing its representation theory origin as
    \bee\label{Beqs}
    &d B+\Omega\star B-B\star \pi(\Omega)-B\star T^{AB} d V_A V_B=0\,.
    \eee
    The last term accounts properly for the $x$-dependence of $V^A$ as the frame field $E^A$ is $(d+\Omega)V$ and $\pi$ knows nothing about $dV$. It is also necessary when checking the integrability, as (\ref{Beqs}) is consistent up to $d\pi$, which is compensated by the last term.

{\bf h.} A standard example to demystify
(\ref{EqLorentzA})-(\ref{EqLorentzB}) is provided by the spin-two,
where $W^{A,B}$ component of $W(T)$ can be decomposed into the
vielbein $W^a$ and spin-connection $W^{a,b}$ and then
(\ref{EqLorentzA}) amounts to the linearized zero-torsion
constraint and the condition that the only nonzero components of the linearized over AdS Riemann two-form
are given by the Weyl tensor $C^{ab,cd}$, c.f. (\ref{OmegaInLorentz})
\bee
&\DL W^a=0,\qquad  &\DL W^{a,b}+\Lambda h^a\wedge
W^b-\Lambda h^b\wedge W^a=h_c\wedge h_d\, C^{ac,bd}
\eee
(\ref{EqLorentzB}) just encodes all derivatives of the Weyl tensor
that are compatible with differential Bianchi identities.

{\bf i.}
(\ref{EqLorentzA}) contains Fronsdal equation (\ref{FronsdalEqs}), which by virtue of (\ref{EqLorentzA}) is imposed on the maximally $V$-parallel component of the spin-$s$ part of $W(T|X)$ field, \bee\label{HSFrametoFronsdal}\fl\qquad\phi^{A(s)}=E^{AM} \, e_M{}^{A(s-1)}\,,\qquad\qquad\quad e^{A(s-1)}=W^{A(s-1),B(s-1)}V_{B(s-1)}\,.\eee where $E\fud{A}{M}$ is the ambient vielbein and $e_M{}^{A(s-1)}$ is a higher-spin vielbein or frame field.

\paragraph{\bf 4. Oscillator realization.}
One can develop a quite effective technic for dealing with
(\ref{EqLorentzA})-(\ref{EqLorentzB}) directly,
\cite{Iazeolla:2008ix}. Unfortunately, it is not known how to
extend this technic beyond the linearized level. Fortunately,
everything can be given by means of oscillator realization, which
does extend to the interaction level, \cite{Vasiliev:2003ev}. One
introduces an $sp(2)$ pair $Y^A_\ga$, $\ga=1,2$, of oscillators,
satisfying, \cite{Vasiliev:2003ev},
\bee
&[Y^A_\ga, Y^B_\gb]_\star=2i\,\eta^{AB}\epsilon_{\ga\gb}\,, \qquad\qquad \epsilon_{\ga\gb}=-\epsilon_{\gb\ga}\,,\qquad \epsilon_{12}=1\,,
\eee
where $\star$ is the Moyal-Weyl $\star$-product
\bee\fl
f(Y)&\star g(Y) =\frac1{(2\pi)^{2(d+2)}}\int\,dU dV \, f(Y+U)
g(Y+V)\, \exp\left(iU^A_\ga V^B_\gb
\epsilon^{\ga\gb}\eta_{AB}\right),
\eee
It is easy to see that
\bee
T^{AB}&=\frac{i}{4}\{Y^A_\ga,\,Y^B_\gb\}_\star\,\epsilon^{\ga\gb}
\eee
satisfy the $so(d,2)$ commutation relations, (\ref{SOCommuRelations}). The
$\star$-product realization makes computations easier than those
with the universal enveloping algebra. That there are only two
species $Y^A_{1,2}$ of oscillators quotients automatically out the
first generator of (\ref{Idealtofactor}). The absence of the
second one, which corresponds to various traces is not granted for
free and it must be factorized by hand via imposing conditions of
type
\bee\label{TraceConstraintsW}\eta^{AB}\frac{\pl^2}{\pl Y^A_\ga \pl Y^B_\gb}...=0\eee that
removes proliferation due to traces. Equation (\ref{EqLorentzB})
can be left unchanged, as it does not depend on the details of
realization while (\ref{EqLorentzA}) now reads
\bee\label{EqLorentzC}
d W+ [\Omega, W]_\star&=\left.-iE^A \wedge E^B\,\epsilon_{\alpha\beta}\frac{\pl^2}{\pl
Y^A_\alpha\pl Y^B_\beta}B(Y)\right|_{Y^A_\ga V_A=0}
\eee

\section{W-propagator, two-point functions}\label{SecWpropagator}
Once the boundary-to-bulk propagator for the Fronsdal field is given (\ref{MikhailovProp}), it is straightforward to determine what the frame field part (\ref{HSFrametoFronsdal}) of $W(T|X)$ is
\bee
e^{A(s-1)}&=\frac{1}{(X Z)^\alpha}\xi^{A(s-1)}\, \xi_N\, dX^N\,,
\eee
where we keep for the moment the weight $\alpha$ free. One has to take the derivatives of (\ref{MikhailovProp}) up to order $s-1$ and to take the traces into account appropriately. The most general ansatz reads (the ambient $X$ now serves also as the compensator)
\bee\fl\label{btobW}
\qquad W^s=&\frac{1}{(X Z)^\alpha}\sum_{0\leq k+2i\leq s-1} A^s_{k,i}\, \WeylD{\xi^{s-1}}{X^{s-1-k-2i}P^kG^i}{} \, \xi_N\, dX^N,\\\fl
&\WeylD{\xi^{s-1}}{X^{s-1-k-2i}P^kG^i}{}\equiv T_{AB(s-1)}\, \xi^{A(s-1)}\, X^{B(s-1-k-2i)} P^{B(k)} G^{BB(i)}\,.
\eee
(\ref{EqLorentzC}) implies that $\DO W$ equals zero almost everywhere in the parameter space, which with the help of (\ref{DerivativesAll}) gives
\bee
A^s_{k,i}&=A^s\frac{(-1)^i \Gamma[\alpha+k-1 ]\, \Gamma\!\left[\frac12{(s+\alpha) }\right]}{i!\, k!\, \Gamma\,[\alpha-1]\, \Gamma\!\left[\frac12(s+\alpha) - i)\right]}\,,
\eee
where $A^s$ reflects the freedom in normalizing any $W^s$ separately. If one now checks whether the trace constraint (\ref{TraceConstraintsW}) is satisfied, one finds up to some nonvanishing function
\bee
\pl^C_\mu \pl_{C\nu}W\sim (2 \Delta+s - \alpha),\qquad\qquad s>2\,,
\eee
which implies that the trace constraint (\ref{TraceConstraintsW}) singles out the
weight of a massless spin-$s$ field, a similar phenomenon was observed in \cite{Bekaert:2012vt}.
One may expect that the connection $W$ with relaxed trace
constraints is still a good starting point for the description of
massive fields too, as $\DO W\approx0$ for any weight $\alpha$.
For $\alpha = 2 \Delta+s$ one finds
\bee
A^s_{k,i}&=A^s\frac{(-1)^i\, \Gamma[s+\Delta ]\, \Gamma[k+s+2 \Delta -1]}{i!\, k!\, \Gamma[s+\Delta -i]\, \Gamma[s+2 \Delta -1]}\,.\label{Wcoefs}
\eee
\paragraph{\bf Weyl tensor.} The terms that do not cancel inside $\DO W$ with (\ref{Wcoefs}) are given by the Weyl tensor part $C^{A(s),B(s)}$ of $B(T|X)$. Actually the form of the Weyl tensor,
\bee\label{SpinSWeyl}
&C^s=\frac{1}{(X Z)^{2\Delta+s}}\sum_{i=0}^{i=[\frac{s}2]} H^s_{i}\, \WeylD{\xi^s}{P^{s-2i}G^i}{}\,,\\
&\WeylD{\xi^s}{P^{s-2i}G^i}{}\equiv T_{AB(s)}\, \xi^{A(s)}\, P^{B(s-2i)} G^{BB(i)}\,,\\
&H^s_i=H^sH^s_i\,,\qquad\qquad H^s_i=\frac{(-)^i s!\, \Gamma\!\left[s+\Delta -\frac{1}{2}-i\right]}{4^i\, i!\, (s-2i)!\, \Gamma\!\left[s+\Delta -\frac{1}{2}\right]}
\eee
is completely fixed up to the overall factor $H^s$ by the requirement
for it to be traceless. It is constructed in terms of variables
that are all tangent to the AdS-hyperboloid. Therefore,
$so(d,2)$-tracelessness implies $so(d,1)$-tracelessness, as it should be. The only
thing to do is to determine the relative normalization of
$H^s$ that does cancel $\DO W$, which gives
\bee
H^s=A^s\frac{\Gamma[2 s+2 \Delta -1]}{(s+1)!\, \Gamma[s+2 \Delta -1]}\,.
\eee
Let us note that for (\ref{EqLorentzC}) to hold for the traceless Weyl tensor the massless fall-off $(2\Delta+s)$ is now mandatory.

\paragraph{\bf Two-point functions.}
Using the prescriptions given in Section \ref{SecGeometry}, it is
easy to see that on approaching the boundary all terms in $W^s$
tend to
\bee
\frac{1}{(X Z)^{2\Delta+s}} T_{AB(s-1)}\, \xi^{A(s-1)}\, X^{B(s-1)} \xi_N\, dX^N
\eee
the latter expression suggests that the role of a polarization vector in the higher-spin theory is played by $\zeta^A=T^{AB}X_B$, which is by definition orthogonal to the would-be soon boundary point $X^A$. Unfortunately it is not null, which manifests the fact that the higher-spin theory is formulated in an extended space of variables, where the conditions like tracelessness do not hold automatically and require a separate and rather cumbersome treatment, \cite{Vasiliev:2003ev, Bekaert:2005vh}. When on the boundary, one recover (\ref{TwoPointCurrent}),
which is a desired correlation function of two spin-$s$ conserved currents.

More easily the two-point functions can be extracted out of the Weyl tensor, which directly tends to
\bee\fl\qquad
\frac{1}{(X Z)^{2\Delta+s}} T_{AB(s)}\, \xi^{A(s)}\, X^{B(s)}=\frac{1}{(X Z)^{2\Delta}} \left(\frac{\zeta_A\, \Pi(X,Z)\fud{A}{B}\,\eta^B}{(X Z)}\right )^s
\eee
which is (\ref{TwoPointCurrent}) again. The phenomenon that the Weyl tensor is relevant for extracting correlation functions of currents was observed in \cite{Giombi:2009wh, Giombi:2010vg}.

\section{B-propagator}\label{SecBpropagator}
The most complicated part is to construct the boundary-to-bulk
propagator for the $B$ field as it contains arbitrarily high
derivatives of the fields. Straightforward approach based on
solving (\ref{EqLorentzB}) seems to be too tedious for spins
greater than zero calling for more refined methods. Nevertheless,
it is first useful to work up scalar case directly.

\subsection{s=0} The case of the scalar field with dimension $2\Delta=d-2$, which
is the lowest component of the bulk higher-spin multiplet, can be
approached rather directly, the most general ansatz being,
\bee\fl\label{SpinZerobtobAnsatz}
B^0=\frac{1}{(XZ)^{\alpha}}F(\nu, u), \quad \nu=\WeylD{P}{X}{}\equiv T^{AB}P_A X_B,\quad u=\WeylD{G}{XX}{}\equiv T^{AB(2)}G_{AA}X_{B(2)}\,,
\eee
where again we keep the fall-off $\alpha$ free. The equations of motion (\ref{EqLorentzB}) lead to
\bee\eqalign{\label{szeroequations}
&\left(-\alpha - N_\nu +\frac12(N_\nu+2N_u+2)\pl_\nu\right)F(\nu,u)=0\,,\\
&\left(-2+\pl_\nu+(N_\nu+2N_u+3)\pl_u\right)F(\nu,u)=0}\,,
\eee
where $N_\nu=\nu\pl_\nu$ and $N_u=u\pl_u$ are the Euler operators. There are two solutions to these equations, the first one, which is simple, and corresponds to the shadow partner of a dimension-$2\Delta$ scalar, which has dimension $2=d-2\Delta$,  reads
\bee\label{ShadowProp}
B^0=\frac{1}{(XZ)^{2}}\exp 2\nu\,.
\eee
The second solution is the one we need. Let us mention that the
shadow solution is simple as it does not depend on variable $u$,
which is of the forth order in $Y^A_\ga$. The latter property is
due to its dimension, which does not involve $d$, so there is no
need in $u$.

To find the second solution it is easier to convert equations (\ref{szeroequations}) into recurrent relations and solve for $F^{k,m}$
\bee
F(\nu,u)=\sum_{k,m}F^{k,m}\nu^k u^m\,,
\eee
the solution being
\bee\label{SpinZeroSolution}
F(\nu,u)=\sum_{k,m} \frac{2^{k+2 m}(-)^m\, \Gamma[\Delta ]\, \Gamma[k+2 \Delta ]}{k!\, m!\, (1+k+2 m)!\, \Gamma[\Delta -m]\, \Gamma[2 \Delta ]}\,\nu^k\, u^m\,.
\eee
Let us make several comments on the solutions obtained.
\begin{enumerate}
  \item If one discards $(1+k+2 m)!$ factor in $F^{k,m}$, (\ref{SpinZeroSolution}),  which, as will become evident soon, appears naturally from the $\star$-product integration, then the solution has a very simple generating function
\bee
\widetilde{F}(\nu,u)=(1-2 \nu)^{-2 \Delta } (1+4 u)^{-1+\Delta }\label{SpinZeroTransformed}
\eee
The additional interfering factorial can be treated with the help
of the Hankel representation for $\Gamma$-function, which in our
case of integer argument reduces to
\bee
\frac{1}{\Gamma(n)}=\frac{1}{2\pi i}\oint_{\mathcal{C}}z^{-n}e^n
dz\,,
\eee
with the closed contour around the origin. The solution is the
transform
\bee
F(\nu,u)=\frac{1}{2\pi i}\oint_{\mathcal{C}}dz\,
\widetilde{F}(\nu/z,u/z^2) z^{-2}e^z \,.\label{SpinZeroBackTrans}
\eee

  \item The form of the solution (\ref{SpinZeroSolution}) depends on the space-time dimension modulo $2$, which is a general phenomenon. Indeed, for $d$ even, i.e. $\Delta$ integer, the Taylor expansion in $u$ stops at $u^{\Delta-1}$ as is seen from (\ref{SpinZeroTransformed}),
  while for $d$ odd, i.e. $\Delta$ half-integer, the solution contains all powers of $u$.
  \item Using a prescription of Section \ref{SecGeometry}
   for extracting two-point functions one observes that both $\nu$ and $u$ tends to zero on approaching the boundary, therefore
      \bee
      \left.\rule{0pt}{14pt}B^0\right|_{boundary}=\frac{1}{(X Z)^{2\Delta}}
      \eee
      as it was expected for $\langle j_0(X)j_0(Z)\rangle$, where $j_0=:\phi(X)\phi(X):$.
  \item As one can readily check both solutions satisfy a twisted analog of the trace constraint
(\ref{TraceConstraintsW}), which has the form
      \bee
      \left(G^{AB}\frac{\pl^2}{\pl Y^A_\ga \pl Y^B_\gb}+Y_A^\ga V^A  Y_B^\gb V^B\right)B^0(Y)=0\,.
      \eee
It is what should have been expected once the scalar is associated
with $B(Y=0)$. Let us note that there is no freedom in choosing
trace factorization condition once the equations of motion are
satisfied and it is stated which component of $B$ is a scalar
field (we have assumed a canonical choice $B(Y=0)$ is a scalar
field).
  \item The exponent $\exp{2\nu}$ that appears in the shadow solution is a distinguished one as it is a
$\star$-algebra projector analogous to the one used in
\cite{Vasiliev:2012vf}, $\exp{2\nu} \star \exp{2\nu}=\exp{2\nu}$,
which is important for going beyond the linearized approximation.
Among other things, it guarantees that the potentially divergent
self-interaction terms for higher-spin fields cancel. One might
think of extracting $\exp{2\nu}$ out of the second solution, the
result being
\bee\fl
 B^0=\frac{\exp{2\nu}}{(XZ)^{2\Delta}}\sum_{k,m}\frac{(-1)^m 2^k \Gamma\left[\Delta-m-\frac{1}{2} \right] \Gamma[2 \Delta -1]}{k! m! (1+k+2 m)! \Gamma\left[\Delta -\frac{1}{2}\right] \Gamma[2 \Delta-k-2 m -1]}\,\nu^k u^m\,.
\eee
\end{enumerate}

There exists a more simple route to (\ref{SpinZeroBackTrans}) to appreciate which we need to go into the details of explicit solutions to (\ref{EqLorentzB}).

\subsection{\bf Twisted-adjoint transformation and the self-similarity method}
Solving the equations of motion to derive $B$-propagator for
a spin greater than zero turns out to be a highly nontrivial
problem. To avoid this problem we propose another approach which
is based on the following propositions:
\begin{enumerate}
\item The lowest component of the unfolded $B$-propagator which is
the Weyl tensor is known and given by (\ref{SpinSWeyl}); \item
Vectors $\xi^{A}$ and $P^A$ the propagator depends upon are
related at different points of $AdS$ space according to
(\ref{TwistedAdjointRot}); \item Solution to twisted-adjoint eq.
(\ref{EqLorentzB}) is pure gauge
\be\label{PureGauge}
B=g^{-1}\star B_0\star \pi(g)
\ee
\end{enumerate}
The idea is as follows. Suppose we know the solution at some point
$X_0^A=V^A$: $B=B(Y_{\alpha}^{A}|\,\refp{V}{\xi}, \refp{V}{P},
(V Z))$, then at an arbitrary point $X$ it amounts to
$B(\overline{Y}_{\alpha}^{A}|\,\refp{X}{\xi}, \refp{X}{P}, (X Z))$,
 where $\overline{Y}_{\alpha}^{A}=\overline{Y}_{\alpha}^{A}(X)$
 receive $X$-dependence. Both solutions are related by twisted-similarity
 transformation (\ref{PureGauge})
 \be\label{SelfSimilar}
 g^{-1}\star B(Y_{\alpha}^{A}|\,\refp{V}{\xi}, \refp{V}{P},
(V Z))\star\pi(g)=B(\overline{Y}_{\alpha}^{A}|\,\refp{X}{\xi},
\refp{X}{P}, (X Z))\,.
 \ee
Note that the same function $B$ enters both sides of
(\ref{SelfSimilar}). Here we assume that $g^{-1}\star dg$ have
all its components in the $so(d,2)$ subalgebra of the higher-spin
algebra, i.e. $g=g(T|X)$ defines some global rotation of AdS,
which takes compensator $V^A$ to $X^A$,
$X^A=\Lambda\fud{A}{B}V^B$, the simplest such a $g(T|x)$ read
\bee
g(T|X)=\exp\left[-2T^{AB}X_A V_B\,(1-X\cdot V)^{-1}\right]\,.
\eee
$\overline{Y}_{\alpha}^{A}$-oscillators should preserve their
commutation relations and, therefore, transform in the adjoint of
$so(d,2)$
\be
\overline{Y}_{\alpha}^{A}=g^{-1}\star Y_{\alpha}^{A}\star
g=\Lambda\fud{A}{B}Y^{B}_{\alpha}\,.
\ee
Equation (\ref{SelfSimilar}) we are going to refer to as the {\it
self-similarity} condition. Restricting (\ref{SelfSimilar}) to
Lorentz sector, i.e. setting $V^{A}Y_{A\alpha}\equiv
X^{A}\overline{Y}_{A\alpha}=0$, one arrives at some integral
equation with only Weyl tensor of (\ref{SpinSWeyl}) remaining on
the {\it r.h.s.} This will eventually allow us to determine $B$-function
completely. Before going into the details, let us consider now
{\it l.h.s} of (\ref{SelfSimilar}). For an arbitrary function $F(Y)$
\bee
\widehat{F}(Y)=g^{-1}(Y) \star F(Y) \star \pi(g(Y))
\eee
simple gaussian integration in terms of $\overline{Y}^A_\ga$ and
$\tau=(XV)$ yields
\bee\fl\label{TwistedAdjointRotation}
&\widehat{F}(\overline{Y}^A_\ga)=\int ds\,dt\,\exp i\left(
\overline{Y}^A_\ga(s^\ga  V_A +t^\ga {X_A})+\tau s^\ga t_\ga
\right)F\left(\overline{Y}^A_\ga+V^A s_\ga+{X^A}t_\ga\right)\,,
\eee
where it is implied that the 'initial data' $F(Y)$ is given at the
point $V$, at which the ambient coordinate $V^A$ coincides with
the compensator field $V^A$. Setting further
$X^{A}\overline{Y}_{A\alpha}=0$, as we are interested to end up
with the Weyl tensor on {\it r.h.s.} of (\ref{SelfSimilar}), one obtains
\bee\label{TwistedAdjointRotationB}
\left.\widehat{F}\right|_{X^{A}\overline{Y}_{A\alpha}=0}=\frac1{\tau^2}\int ds\,dt\, F\left(\hat{Y}^A_\ga+V^A s_\ga+\frac{1}{\tau}X^A t_\ga\right)\, e^{it_\ga s^\ga },\\
\hat{Y}^A_\ga=\Pi(V,X)\fud{A}{B}\overline{Y}^{B}_\ga\,,\qquad\qquad
\Pi_{A}{}^{C}\Pi_{CB}=\Pi_{AB}\,,
\eee
where $\Pi(V,X)\fud{A}{B}$ (see Table \ref{TblGeometric}) now
comes as the projector to Lorentz directions. This means, in
particular, that $\hat{Y}^{A}_{\alpha}V_A=0$. Another useful in
what follows observation is
\be\label{YetUseful}
\refp{X}{\xi_A}\,\Pi(V,X)\fud{A}{B}=\refp{X}\xi_{B}\,,\qquad
\refp{X}{P_A}\,\Pi(V,X)\fud{A}{B}=\refp{X}P_{B}\,.
\ee
Having simple propagator for shadow scalar field
(\ref{ShadowProp}) at hand, let us illustrate how self-similarity
equation (\ref{SelfSimilar}) indeed performs the desired
transformation
\bee
Y\rightarrow\overline{Y}\qquad V\rightarrow X \qquad
\refp{V}{P}\rightarrow\refp{X}{P}\qquad
\refp{V}{\xi}\rightarrow\refp{X}{\xi}\,,
\eee
where $\refp{X,V}{P}$ and $\refp{X,V}{\xi}$ are the 'wave-vector'
and polarization at points $X$ and $V$ while the boundary point
$Z$ and polarization vector $\eta$ are kept fixed, these were
defined in (\ref{TwistedAdjointRot}). To do so we take the propagator for the
shadow field (\ref{ShadowProp}), replace $X$ by $V$ and apply
(\ref{TwistedAdjointRotation}), one then finds
\bee\fl
&g^{-1}\star B^0\left(Y| \refp{V}{P}, (V Z)\right)\star
\pi(g)=B^0\left(\overline{Y}| \refp{X}{P}, (X Z)\right)\,,\\\fl\qquad
g^{-1}\star& \frac1{(V Z)^2} \exp{2[\refp{V}{P_A}\, T^{AB}
V_B]}\star \pi(g)=\frac1{(X Z)^2} \exp{2[\refp{X}{P_A}\,
\overline{T}^{AB}X_B]}\,.
\eee
The latter means that the twisted-adjoint rotation transforms the
boundary-to-bulk propagator from $Z$ to $V$ to the one from $Z$ to
$X$. In particular, the integration produces a prefactor that
changes $(V Z)^{-2}$ to $(X Z)^{-2}$.

For $s>0$ analysis we need to elaborate
(\ref{TwistedAdjointRotationB}) a bit further. Since all of the
functions depend only on $T^{AB}$, it is useful to look at what
these transform into. Let us expand $F(Y)$ in terms of $T^{AB}$
\bee
F(Y)=F(T)=\sum_N T^{AB(N)}\, C_{A(N)|B(N)}\,,
\eee
where $C_{A(N)|B(N)}$ are symmetric in each group of indices.
Using (\ref{TwistedAdjointRotation}) and the orthogonality
condition
\bee\fl
\int ds\,dt\,  (s^\ga t_\ga)^m (s^\ga \xi_\ga)^{k}(t^\gb
\eta_\gb)^{l} \exp i\left(s^\ga t_\ga
\right)=\delta_{k,l}\frac{i^{k+m}(k+m+1)!}{k+1}
(\xi^\ga\eta_\ga)^k\,,
\eee
where $\xi$ and $\eta$ are auxiliary spinors, the term-wise result
is
\bee\eqalign{\label{Ftwistedrotated}
\left.\widehat{F}\right|_{X^{A}\overline{Y}_{A\alpha}=0}=\frac1{\tau^2}\sum_{N=0}^{\infty}\sum_{s=0}^{N} D^s_N C_N^s\, \hat{T}^{AB(s)}\Theta^{AB(N-s)}\,C_{A(N)|B(N)}\,,\\
D^s_N=\frac{(-)^{N+s}(N+1)!}{2^{N-s}(s+1)!}\,,\qquad\qquad C_N^s=\frac{N!}{(N-s)!s!}\,,}
\eee
where we introduced the notation
\be
\Theta^{A,B}=\frac1{\tau} (X^A V^B-X^B V^A)\,.
\ee
The idea of obtaining the spin-$s$ propagator is to
reconstruct the unknown function $B$ from the Weyl tensor
(\ref{SpinSWeyl}) using self-similarity equation
(\ref{SelfSimilar}) at $V^{A}Y_{A\alpha}\equiv
X^A\overline{Y}_{A\alpha}=0$. It is instructive to first consider
spin-zero case separately before we proceed with arbitrary spin
propagator.

\subsection{\bf Self-similarity for spin zero.}
Let us apply the elaborated method to the simplest case of spin
zero. Following the logic of the previous section, suppose that we
already know the boundary-to-bulk propagator at a point $V$, where
by a coincidence the ambient coordinate $V^A$ equals the
compensator field $V^A$. It is given by an expression similar to
(\ref{SpinZerobtobAnsatz})
\bee\fl\label{SpinZerobtobAnsatzV}
\refp{V}{B}=\frac{1}{(V Z)^{2\Delta}}F(\nu, u)\,, \qquad\qquad
\nu=\refp{V}{P_A}\,T^{AB}V_B\,,\qquad
u=\refp{V}{G_{AC}}\,T^{AB}T^{CD}V_BV_D\,.
\eee
When twisted-adjointly rotated with $g(Y)$ that takes $V$ to $X$
the propagator must coincide with  (\ref{SpinZerobtobAnsatz}). In
particular, $\refp{X}{B}$ at $\overline{Y}^A_\ga=0$ must
be\footnote{For $s=0$ setting $X^A\overline{Y}_{A\alpha}=0$ is
equivalent to $\overline{Y}_{A\alpha}=0$} the 'two-point function'
$(X Z)^{-2\Delta}$. With (\ref{TwistedAdjointRotationB}) one
finds, $\tau=(XV)$
\bee
&\refp{X}{B}|_{\overline{Y}=0}=\frac1{\tau^2}\int ds\,dt\,\exp
i\left( s^\ga t_\ga \right)\left.\refp{V}{B}\left(T^{AB}\, s^\ga
t_\ga\right)\rule{0pt}{12pt}\right|_{T^{AB}=\Theta^{AB}}\,.
\eee
Which leads us to a transform
\bee\fl\label{trans}
\int ds\,dt\,\exp i\left( s^\ga t_\ga \right) f(s^\ga t_\ga\,
x)=\sum_k f_k\, x^k \, i^k(k+1)!\qquad \qquad f(x)=\sum_k f_k\,x^k
\eee
that brings an additional factorial due to $\int ds\,dt\, (s^\ga
t_\ga)^k\exp i\left( s^\ga t_\ga \right)=i^k(k+1)!$. The factorial
just counts the number of $T^{AB}$. Let us note, that the
transformation (\ref{trans}) is to some extent reminiscent to the
one used in \cite{Prokushkin:1998vn} in different context. For a function of
two variables (\ref{SpinZerobtobAnsatzV}), the transformation
reads
\bee
\refp{X}{B^0}|_{\overline{Y}=0}=\frac{1}{\tau^2}\frac{1}{(V Z)^{2\Delta}} \left. \widetilde{F}(\nu,u)\right|_{T^{AB}=\Theta^{AB}}\,,\\
\widetilde{F}(x,y)=\sum_{k,m} F_{k,m}\,x^k
y^m\,\frac{(k+2m+1)!(-)^k}{2^{k+2m}}\,,
\eee
with the inverse map given by
\be\label{InverseTrans}
F(x,y)=\frac{1}{2\pi i}\oint\,dz\,\frac{e^z}{z^2}
\widetilde{F}\left(-\frac{2x}{z}, \frac{4y}{z^2}\right)\,.
\ee
Using convenient variables $\sigma=(X Z)/(V Z)$ and
$\tau=(X V)$, we have
\bee\fl\label{Variables}
\left.\nu\rule{0pt}{12pt}\right|_{T^{AB}=\Theta^{AB}}=-1-\frac{(X Z)}{(V Z)(X V)}=-1-\sigma\tau^{-1}\qquad\qquad
\left. u\rule{0pt}{12pt}\rule{0pt}{12pt}\right|_{T^{AB}=\Theta^{AB}}=1-\tau^{-2}
\eee
Amazingly, the condition for $\refp{X}{B^0(\overline{Y}=0)}$ to
have the correct behaviour is sufficient to fix the function of
two variables, thus determining the propagator completely without
solving any differential equations at all! Indeed,
\bee\fl
\frac1{(X
Z)^{2\Delta}}=\refp{X}{B^0(\overline{Y}=0)}=g^{-1}\star
\refp{V}{B^0(Y)}\star
\pi(g)|_{\overline{Y}=0}=\frac{\widetilde{F}(-1-\sigma\tau^{-1},1-\tau^{-2})}{\tau^2(V
Z)^{2\Delta}}
\eee
which immediately gives
$\widetilde{F}(-1-\sigma\tau^{-1},1-\tau^{-2})=\tau^2
\sigma^{-2\Delta}$. The straightforward inverse transform
$\widetilde{F}(\nu,u)\rightarrow F(\nu,u)$ gives
(\ref{SpinZeroSolution}), which has been obtained by a direct
solving of the field equations. The simple form of the generating
function (\ref{SpinZeroTransformed}) comes now without surprise as
well as the dependence of the solution on $d\,\mbox{mod}\,2$.

A form of the scalar propagator which explicitly contains
star-product projector might be of use for application. To obtain
it one should take
\be
\refp{V}{B}=\frac{1}{(V Z)^{2\Delta}}e^{2\nu}F(\nu,
\omega)\,,\qquad \omega=u-\nu^2\,.
\ee
Repeating the above procedure one arrives at
\be\label{ScalarIntegral}
B=\frac{\exp{2\nu}}{(X Z)^{2\Delta}}\,\frac{1}{2\pi i}\oint\,dz\,
\frac{e^z}{z^2}\left(1+\frac{4\nu}{z}-\frac{4\omega}{z^2}\right)^{\Delta-1}\,.
\ee
The residue in (\ref{ScalarIntegral}) is some polynomial of a
degree $\Delta-1$ for integer $\Delta$ and infinite series for
$\Delta$ half-integer. Particularly, the residue equals just 1 for
$\Delta=1$. As we see, this method is simple yet effective and
will be applied to a general case of spin-$s$ field below.

\subsection{Any s}
Again, instead of solving (\ref{EqLorentzC}) directly we will use
the self-similarity of the propagator (\ref{SelfSimilar}), where
$g$ rotates $V$ to $X$. This means that within the ambient
approach the propagator looks the same at any point. Bearing in
mind the spin-zero case, we may have a look only at the Weyl tensor
at point $X$ as viewed from point $V$ via (\ref{SelfSimilar}), i.e., try to solve for
\bee\fl\label{BSelfEquation}
(\ref{SpinSWeyl})=\left.B(\refp{X}{P},\refp{X}{\xi},\refp{X}{G},(X Z))\right|_{\overline{Y}^A_\ga X_A=0}=g^{-1}\star\left.B(\refp{X}{P},\refp{X}{\xi},\refp{X}{G},(V Z))\star \pi(g)\right|_{\overline{Y}^A_\ga X_A=0}
\eee
with the hope that it again determines the dependence on all
variables, and it does.

First of all one faces the problem of parameterizing various
structures that can appear in the $B$-field. If
factorized it amounts to 10 variables, some of them satisfying
quadratic relations, which makes a direct solving somewhat
complicated. All the descendants of order $N-s$ of the spin-$s$
Weyl tensor can be parameterized as
\bee\fl\nonumber
\WeylD{\xi^s P^{k+q}G^{m-q}}{P^{s-2n-q}G^nV^{k+2m}}{G^q}\equiv
T^{AB(N)}\xi_{A(s)}P_{A(k+q)}G_{AA(m-q)}G_{AB(q)}P_{B(s-2n-q)}G_{BB(n)}V_{B(N-s)}\\
\nonumber 0\leq 2n+q\leq s,\qquad\qquad k+2m=N-s,\qquad\qquad q\leq
m\,,\eee which gives for the $B$ field in the spin-$s$ sector, $B^s$,
\bee\label{SpinSAnsatz}
B^s=\sum_{N}\sum_{k+2m=N}\sum_{0\leq 2n+q\leq
s}F^{k,m}_{n,q}\WeylD{\xi^s
P^{k+q}G^{m-q}}{P^{s-2n-q}G^nV^{k+2m}}{G^q}
\eee
That it is a complete basis can be seen either by evaluating the
tensor product $\xi^s \otimes V^N\otimes G^{m+n}\otimes
P^{s+k-2n}$ or by noting that given any arrangement of
$\xi,P,G,V$'s inside a tensor having the symmetry properties of a
rectangular two row Young diagram one can always push all $\xi$'s
to the first group of symmetrized indices and all $V$'s to the
second. After that there is no freedom left in rearranging the
indices while preserving $\xi$'s and $V$'s, so the rest of $P$'s
and $G$'s can appear in any combination, as they do above.

At the new point $X$, the role of the compensator field is played
by $X$ itself. Thus, one sets $\overline{Y}^A_\ga X_A=0$ and
notices that $\hat{T}^{AB}$ becomes orthogonal to the old
compensator $V$, $\hat{T}^{AB} V_B=0$, while $\Theta^{A,B}$ has no
components into the Lorentz subspace, i.e. contraction of
$\Theta^{A,B}$ with any Lorentz tensor vanishes identically.
Therefore (\ref{SpinSAnsatz}) can be contracted with
$\hat{T}^{AB(s)} \Theta^{AB(N-s)}$ only and
(\ref{Ftwistedrotated}) simplifies to
\bee\label{FtwistedrotatedB}
\widehat{F}(\hat{T}^{AB})=\frac1{\tau^2}\sum_{N=0}^{\infty}
D^s_N C^s_N\, \hat{T}^{AB(s)}\Theta^{AB(N-s)}\,C_{A(N)|B(N)}
\eee
Now one just needs 1) to express the new Weyl tensor at point $X$,
(\ref{SpinSWeyl}), in terms of the old variables $\refp{V}{P}$,
$\refp{V}{\xi}$, $\refp{V}{G}$, the relevant transformations
having been already given in (\ref{TwistedAdjointRot}); 2)
substitute (\ref{SpinSAnsatz}) into (\ref{FtwistedrotatedB}) and
expand. Matching various structures at both sides of (\ref{BSelfEquation}) one finds
all the $F^{k,m}_{n,q}$.

{\bf1.} There are two structures that contribute to the Weyl
tensor (\ref{SpinSWeyl}), with the help of
(\ref{TwistedAdjointRot}) and (\ref{YetUseful}) one derives
\bee
C^s=\frac1{(X Z)^{2\Delta+s}}\sum_i B^s_i \WeylD{\refp{X}{\xi}}{\refp{X}{P}}{}^{s-2i}\WeylD{\refp{X}{\xi}\refp{X}{\xi}}{\refp{X}{G}}{}^i\\
\WeylD{\refp{X}{\xi}}{\refp{X}{P}}{}=\frac1{\sigma}\WeylD{{\xi}}{{P}}{}-\Upsilon\qquad\qquad
\Upsilon=\tau \WeylD{{\xi}}{{V}}{G}+\frac{\tau^2}{\sigma}\WeylD{{P}}{V}{G}\WeylD{{\xi}}{{V}}{}\\
\WeylD{\refp{X}{\xi}\refp{X}{\xi}}{\refp{X}{G}}{}=\WeylD{{\xi\xi}}{{G}}{}+\frac{2\tau}{\sigma}\WeylD{{\xi
P}}{{G}}{}\WeylD{{\xi}}{{V}}{}+\frac{\tau^2}{\sigma^2}\WeylD{{P\,
P}}{{G}}{}\WeylD{{\xi}}{{V}}{}^2+\Upsilon^2
\eee
where all variables on {\it r.h.s.} refer to the point $V$, so the
superscript $V$ is dropped. In order to determine all the
$F^{k,m}_{n,q}$ one does not need to expand the Weyl tensor in
full, matching some signature terms is sufficient. We would like
to look at
\bee
\mathbf{S}_{n,q}=
\WeylD{{\xi\xi}}{{G}}{}^n\WeylD{{\xi}}{{P}}{}^{s-2n-q}\WeylD{{\xi}}{{V}}{G}^q\,,
\eee
for which one finds
\bee\fl
C^s=\frac1{(X Z)^{2\Delta+s}}\sum_{0\leq 2n+q\leq s} H^s L^s_{n,q}\, \mathbf{S}_{n,q}\qquad\qquad
L^s_{n,q}=\sum_{i=n}^{i=n+[\frac{q}{2}]} H^s_i C_{s-2i}^{2n+q-2i}C_i^n\\
L^s_{n,q}=\frac{s!\,(-)^n\,\Gamma[\Delta+s-n-[\frac{q}2]-\frac12]\,\Gamma[\Delta+s-n-[\frac{q+1}2]]}
{4^n\,(s-2n-q)!\,n!\,q!\,\Gamma[\Delta+s-\frac12]\,\Gamma[\Delta+s-n-q]}
\eee
{\bf2.} Simple combinatorics with (\ref{SpinSAnsatz}) results in
\bee\fl\eqalign{\nonumber
\WeylD{\xi^s P^{k+q}G^{m-q}}{P^{s-2n-q}G^nV^{k+2m}}{G^q}=\,&\mathbf{S}_{n,q}\, (C_N^s)^{-1}\WeylD{{P}}{{V}}{}^{k+q}
\WeylD{{G}}{{VV}}{}^{m-q}+...
}
\eee
where $...$ denotes the terms with other rearrangements of $\xi$. It is useful do define $\widetilde{F}_{n,q}$ as
\bee
\widetilde{{F}}_{n,q}(x,y)=\sum_N\sum_{k+2m=N}  D^s_N F^{k,m} x^{k+q} y^{m-q}
\eee
where $x$ and $y$ are the same as for spin-zero case, (\ref{Variables}), given by
\bee
x=\left.\WeylD{{P}}{{V}}{}\right|_{T=\Theta}=-1-\sigma\tau^{-1}\,,\qquad\qquad
y=\left.\WeylD{{G}}{{VV}}{}\right|_{T=\Theta}=1-\tau^{-2}
\eee
and introduce
$\widehat{F}_{n,q}(\tau,\sigma)=\widetilde{F}_{n,q}(-1-\sigma\tau^{-1},1-\tau^{-2})$.
Matching $S_{n,q}$ terms on both sides one directly finds, $\vartheta=\Delta+s-n$,
\bee
\widehat{F}_{n,q}=H^s L^s_{n,q}(-)^q  \tau^{q+2} \sigma^{-2\vartheta+q}
\eee
and performing an inverse transform results in,
\bee\fl\label{Bcoefs}
F^{k,m}_{n,q}=H^s L^s_{n,q}\frac{(-)^m 2^{k+2m}(s+1)!}{(s+k+2m+1)!(m-q)!(k+q)!}\,
\frac{\Gamma[\vartheta-m]\,\Gamma[2\vartheta+k]}{\Gamma[\vartheta-q]\,\Gamma[2\vartheta-q]}
\eee
One might worry about the subleading terms that appear on the both
sides, these should match automatically, the explicit computations
being more involved though. To have an additional control over the
computations we have explicitly checked that the first subleading
terms in which one of the $\xi$'s is rearranged in a different way
do match. This check is equivalent to certain nontrivial
differential equations that involve functions $F_{n,q}$ (a
derivative of $F_{n,q}$ brings a factor that is related to the
conformal weight $2\Delta+s$) as well as the coefficients $B^s_i$
that was determined independently from the trace conditions on
Weyl tensor. Of course for $s=0$, for which also $n=q=0$, one
finds a complete agreement with the earlier computations.

\section{Conclusions}\label{SecConclusions}
The boundary-to-bulk propagators for higher-spin master one-form
$W$ (\ref{btobW}), (\ref{Wcoefs}) and zero-form $B$ (\ref{SpinSAnsatz}), (\ref{Bcoefs}) have been constructed in arbitrary dimension
for arbitrary integer spin. The pursuit has forced us to elaborate
an appropriate formalism to tackle this problem as the direct
approach based on solving the e.o.m appeared to be too involved.
The developed self-similarity method is essentially based on the
mixture of the ambient and the unfolded machineries along with the
star-product integration. In fact, the unfolding approach
reduces the problem to purely algebraic. Having the spin-$s$ Weyl
tensor propagator, one defines the generating function that
reproduces its all on-shell derivatives through a simple integral
transformation. The very existence of such an approach was
possible for a number of reasons. First, within the ambient
formalism the propagator is fixed once it is known at a single
point. Second, its pure gauge form within the unfolded formalism
allows one to relate the solution at different points using
similarity transformation, realized as large twisted-adjoint
rotations via $\star$-product. Finally, the latter being presented
in terms of the star-product transformation can be explicitly
evaluated resulting in the above mentioned generating function.

The results obtained in the paper may have at least two
applications. First, one may compute three-point functions in
$d$-dimensional higher-spin theory \cite{Vasiliev:2003ev},
demonstrating how three-point functions of currents built of a
free scalar emerge if the arguments of \cite{Maldacena:2011jn} can
be extended to higher dimensions. As an alternative approach one
may think of putting the higher-spin theory to the boundary
directly using \cite{Vasiliev:2012vf} or developing AdS/CFT
technic with the recently proposed action principle
\cite{Boulanger:2011dd}. Second, according to the general recipe
of AdS/CFT three-point functions of conserved currents $\langle
j_{s_1}j_{s_2}j_{s_3}\rangle$ are in one-to-one correspondence
with various cubic vertices one can construct for massless fields
with spins $s_1$, $s_2$, $s_3$, \cite{Costa:2011mg}. There is a
number of such vertices, \cite{Metsaev:1993ap, Metsaev:2005ar},
which results in a number of independent structures that can
contribute to $\langle j_{s_1}j_{s_2}j_{s_3}\rangle$, found
recently in \cite{Zhiboedov:2012bm}. One may plug the propagators
into the cubic vertices, considered in \cite{Vasilev:2011xf}, to
get the AdS part of the cubic vertices/three-point functions
dictionary.

\ack We would like to thank K.B.Alkalaev, N.Boulanger,
M.A.Grigoriev, R.R.Metsaev, Ph.Spindel, Per Sundell, V.N.Zaikin
for many valuable discussions and especially M.A.Vasiliev for
reading the manuscript and making valuable comments. E.S. thanks I.H.E.S. for hospitality.
The work of E.S. was supported by the Alexander von Humboldt
Foundation. The work of V.D. was supported in part by the grant of
the Dynasty Foundation. The work of E.S. and V.D. was supported in
part by RFBR grant No.11-02-00814 and Russian President grant No.
5638.

\section*{References}
\providecommand{\href}[2]{#2}\begingroup\raggedright\endgroup


\begin{thebibliography}{10}

\bibitem{Maldacena:1997re}
J.~M. Maldacena, {\it {The large N limit of superconformal field theories and
  supergravity}},  {\em Adv. Theor. Math. Phys.} {\bf 2} (1998) 231--252
  [\href{http://arXiv.org/abs/hep-th/9711200}{{\tt hep-th/9711200}}].

\bibitem{Gubser:1998bc}
S.~S. Gubser, I.~R. Klebanov and A.~M. Polyakov, {\it {Gauge theory correlators
  from non-critical string theory}},  {\em Phys. Lett.} {\bf B428} (1998)
  105--114 [\href{http://arXiv.org/abs/hep-th/9802109}{{\tt hep-th/9802109}}].

\bibitem{Witten:1998qj}
E.~Witten, {\it {Anti-de Sitter space and holography}},  {\em Adv. Theor. Math.
  Phys.} {\bf 2} (1998) 253--291
  [\href{http://arXiv.org/abs/hep-th/9802150}{{\tt hep-th/9802150}}].

\bibitem{Vasiliev:1990en}
M.~A. Vasiliev, {\it Consistent equation for interacting gauge fields of all
  spins in (3+1)-dimensions},  {\em Phys. Lett.} {\bf B243} (1990) 378--382.

\bibitem{Vasiliev:1997dq}
M.~A. Vasiliev, {\it Deformed oscillator algebras and higher-spin gauge
  interactions of matter fields in 2+1 dimensions},
  \href{http://arXiv.org/abs/hep-th/9712246}{{\tt hep-th/9712246}}.

\bibitem{Vasiliev:1997ak}
M.~A. Vasiliev, {\it Higher-spin-matter gauge interactions in 2+1 dimensions},
  {\em Nucl. Phys. Proc. Suppl.} {\bf 56B} (1997) 241--252.

\bibitem{Prokushkin:1998vn}
S.~Prokushkin and M.~A. Vasiliev, {\it 3d higher-spin gauge theories with
  matter},  \href{http://arXiv.org/abs/hep-th/9812242}{{\tt hep-th/9812242}}.

\bibitem{Prokushkin:1998bq}
S.~F. Prokushkin and M.~A. Vasiliev, {\it Higher-spin gauge interactions for
  massive matter fields in 3d ads space-time},  {\em Nucl. Phys.} {\bf B545}
  (1999) 385 [\href{http://arXiv.org/abs/hep-th/9806236}{{\tt
  hep-th/9806236}}].

\bibitem{Vasiliev:2003ev}
M.~A. Vasiliev, {\it Nonlinear equations for symmetric massless higher spin
  fields in (a)ds(d)},  {\em Phys. Lett.} {\bf B567} (2003) 139--151
  [\href{http://arXiv.org/abs/hep-th/0304049}{{\tt hep-th/0304049}}].

\bibitem{Vasiliev:1995dn}
M.~A. Vasiliev, {\it Higher-spin gauge theories in four, three and two
  dimensions},  {\em Int. J. Mod. Phys.} {\bf D5} (1996) 763--797
  [\href{http://arXiv.org/abs/hep-th/9611024}{{\tt hep-th/9611024}}].

\bibitem{Vasiliev:1999ba}
M.~A. Vasiliev, {\it Higher spin gauge theories: Star-product and ads space},
  \href{http://arXiv.org/abs/hep-th/9910096}{{\tt hep-th/9910096}}.

\bibitem{Bekaert:2005vh}
X.~Bekaert, S.~Cnockaert, C.~Iazeolla and M.~Vasiliev, {\it {Nonlinear higher
  spin theories in various dimensions}},
  \href{http://arXiv.org/abs/hep-th/0503128}{{\tt hep-th/0503128}}.

\bibitem{Sundborg:2000wp}
B.~Sundborg, {\it {Stringy gravity, interacting tensionless strings and
  massless higher spins}},  {\em Nucl. Phys. Proc. Suppl.} {\bf 102} (2001)
  113--119 [\href{http://arXiv.org/abs/hep-th/0103247}{{\tt hep-th/0103247}}].

\bibitem{Sezgin:2002rt}
E.~Sezgin and P.~Sundell, {\it {Massless higher spins and holography}},  {\em
  Nucl.Phys.} {\bf B644} (2002) 303--370
  [\href{http://arXiv.org/abs/hep-th/0205131}{{\tt hep-th/0205131}}].

\bibitem{Klebanov:2002ja}
I.~R. Klebanov and A.~M. Polyakov, {\it {AdS dual of the critical O(N) vector
  model}},  {\em Phys. Lett.} {\bf B550} (2002) 213--219
  [\href{http://arXiv.org/abs/hep-th/0210114}{{\tt hep-th/0210114}}].

\bibitem{Chang:2012kt}
C.-M. Chang, S.~Minwalla, T.~Sharma and X.~Yin, {\it {ABJ Triality: from Higher
  Spin Fields to Strings}},  \href{http://arXiv.org/abs/1207.4485}{{\tt
  1207.4485}}.

\bibitem{Konstein:2000bi}
S.~Konstein, M.~Vasiliev and V.~Zaikin, {\it {Conformal higher spin currents in
  any dimension and AdS / CFT correspondence}},  {\em JHEP} {\bf 0012} (2000)
  018 [\href{http://arXiv.org/abs/hep-th/0010239}{{\tt hep-th/0010239}}].

\bibitem{Leonhardt:2002sn}
T.~Leonhardt, A.~Meziane and W.~Ruhl, {\it {On the proposed AdS dual of the
  critical O(N) sigma model for any dimension}},  {\em Phys.Lett.} {\bf B555}
  (2003) 271--278 [\href{http://arXiv.org/abs/hep-th/0211092}{{\tt
  hep-th/0211092}}].

\bibitem{Sezgin:2003pt}
E.~Sezgin and P.~Sundell, {\it {Holography in 4D (super) higher spin theories
  and a test via cubic scalar couplings}},  {\em JHEP} {\bf 0507} (2005) 044
  [\href{http://arXiv.org/abs/hep-th/0305040}{{\tt hep-th/0305040}}].

\bibitem{Leigh:2003gk}
R.~G. Leigh and A.~C. Petkou, {\it {Holography of the N=1 higher spin theory on
  AdS(4)}},  {\em JHEP} {\bf 0306} (2003) 011
  [\href{http://arXiv.org/abs/hep-th/0304217}{{\tt hep-th/0304217}}].

\bibitem{Diaz:2006nm}
D.~E. Diaz and H.~Dorn, {\it {On the AdS higher spin / O(N) vector model
  correspondence: Degeneracy of the holographic image}},  {\em JHEP} {\bf 0607}
  (2006) 022 [\href{http://arXiv.org/abs/hep-th/0603084}{{\tt
  hep-th/0603084}}].

\bibitem{Maldacena:2011jn}
J.~Maldacena and A.~Zhiboedov, {\it {Constraining Conformal Field Theories with
  A Higher Spin Symmetry}},  \href{http://arXiv.org/abs/1112.1016}{{\tt
  1112.1016}}.

\bibitem{Bekaert:2012ux}
X.~Bekaert, E.~Joung and J.~Mourad, {\it {Comments on higher-spin holography}},
   \href{http://arXiv.org/abs/1202.0543}{{\tt 1202.0543}}.

\bibitem{Giombi:2009wh}
S.~Giombi and X.~Yin, {\it {Higher Spin Gauge Theory and Holography: The
  Three-Point Functions}},  {\em JHEP} {\bf 1009} (2010) 115
  [\href{http://arXiv.org/abs/0912.3462}{{\tt 0912.3462}}].

\bibitem{Giombi:2010vg}
S.~Giombi and X.~Yin, {\it {Higher Spins in AdS and Twistorial Holography}},
  {\em JHEP} {\bf 1104} (2011) 086 [\href{http://arXiv.org/abs/1004.3736}{{\tt
  1004.3736}}].

\bibitem{Campoleoni:2010zq}
A.~Campoleoni, S.~Fredenhagen, S.~Pfenninger and S.~Theisen, {\it {Asymptotic
  symmetries of three-dimensional gravity coupled to higher-spin fields}},
  {\em JHEP} {\bf 1011} (2010) 007 [\href{http://arXiv.org/abs/1008.4744}{{\tt
  1008.4744}}].

\bibitem{Henneaux:2010xg}
M.~Henneaux and S.-J. Rey, {\it {Nonlinear $W_{infinity}$ as Asymptotic
  Symmetry of Three-Dimensional Higher Spin Anti-de Sitter Gravity}},  {\em
  JHEP} {\bf 1012} (2010) 007 [\href{http://arXiv.org/abs/1008.4579}{{\tt
  1008.4579}}].

\bibitem{Vasiliev:2001wa}
M.~A. Vasiliev, {\it Cubic interactions of bosonic higher spin gauge fields in
  ads(5)},  {\em Nucl. Phys.} {\bf B616} (2001) 106--162
  [\href{http://arXiv.org/abs/hep-th/0106200}{{\tt hep-th/0106200}}].

\bibitem{Eastwood:2002su}
M.~G. Eastwood, {\it {Higher symmetries of the Laplacian}},  {\em Annals Math.}
  {\bf 161} (2005) 1645--1665 [\href{http://arXiv.org/abs/hep-th/0206233}{{\tt
  hep-th/0206233}}].

\bibitem{Fradkin:1985am}
E.~Fradkin and A.~A. Tseytlin, {\it {CONFORMAL SUPERGRAVITY}},  {\em
  Phys.Rept.} {\bf 119} (1985) 233--362.

\bibitem{Vasiliev:2009ck}
M.~Vasiliev, {\it {Bosonic conformal higher-spin fields of any symmetry}},
  {\em Nucl.Phys.} {\bf B829} (2010) 176--224
  [\href{http://arXiv.org/abs/0909.5226}{{\tt 0909.5226}}].

\bibitem{Dirac:1935zz}
P.~Dirac, {\it {The Electron Wave Equation in De-Sitter Space}},  {\em Annals
  Math.} {\bf 36} (1935) 657--669.

\bibitem{Dirac:1936fq}
P.~A. Dirac, {\it {Wave equations in conformal space}},  {\em Annals Math.}
  {\bf 37} (1936) 429--442.

\bibitem{Fronsdal:1978vb}
C.~Fronsdal, {\it {Singletons and Massless, Integral Spin Fields on de Sitter
  Space (Elementary Particles in a Curved Space. 7.}},  {\em Phys.Rev.} {\bf
  D20} (1979) 848--856.

\bibitem{Eastwood:1991}
M.~Eastwood and C.~Graham, {\it {Invariants of conformal densities}},  {\em
  Duke Math. Jour.} {\bf 63} (1991) 633--671.

\bibitem{Metsaev:1995re}
R.~Metsaev, {\it {Massless mixed symmetry bosonic free fields in d-dimensional
  anti-de Sitter space-time}},  {\em Phys.Lett.} {\bf B354} (1995) 78--84.

\bibitem{Metsaev:1997nj}
R.~Metsaev, {\it {Arbitrary spin massless bosonic fields in d-dimensional
  anti-de Sitter space}},  {\em Lect.Notes Phys.} {\bf 524} (1997) 331--340
  [\href{http://arXiv.org/abs/hep-th/9810231}{{\tt hep-th/9810231}}].

\bibitem{Preitschopf:1998ei}
C.~Preitschopf and M.~A. Vasiliev, {\it {Conformal field theory in conformal
  space}},  {\em Nucl.Phys.} {\bf B549} (1999) 450--480
  [\href{http://arXiv.org/abs/hep-th/9812113}{{\tt hep-th/9812113}}].

\bibitem{Bars:1998ph}
I.~Bars, C.~Deliduman and O.~Andreev, {\it {Gauged duality, conformal symmetry
  and space-time with two times}},  {\em Phys.Rev.} {\bf D58} (1998) 066004
  [\href{http://arXiv.org/abs/hep-th/9803188}{{\tt hep-th/9803188}}].

\bibitem{Barnich:2006pc}
G.~Barnich and M.~Grigoriev, {\it {Parent form for higher spin fields on
  anti-de Sitter space}},  {\em JHEP} {\bf 0608} (2006) 013
  [\href{http://arXiv.org/abs/hep-th/0602166}{{\tt hep-th/0602166}}].

\bibitem{Boulanger:2008up}
N.~Boulanger, C.~Iazeolla and P.~Sundell, {\it {Unfolding Mixed-Symmetry Fields
  in AdS and the BMV Conjecture: I. General Formalism}},  {\em JHEP} {\bf 0907}
  (2009) 013 [\href{http://arXiv.org/abs/0812.3615}{{\tt 0812.3615}}].

\bibitem{Bekaert:2009fg}
X.~Bekaert and M.~Grigoriev, {\it {Manifestly conformal descriptions and higher
  symmetries of bosonic singletons}},  {\em SIGMA} {\bf 6} (2010) 038
  [\href{http://arXiv.org/abs/0907.3195}{{\tt 0907.3195}}].

\bibitem{Costa:2011mg}
M.~S. Costa, J.~Penedones, D.~Poland and S.~Rychkov, {\it {Spinning Conformal
  Correlators}},  {\em JHEP} {\bf 1111} (2011) 071
  [\href{http://arXiv.org/abs/1107.3554}{{\tt 1107.3554}}].

\bibitem{SimmonsDuffin:2012uy}
D.~Simmons-Duffin, {\it {Projectors, Shadows, and Conformal Blocks}},
  \href{http://arXiv.org/abs/1204.3894}{{\tt 1204.3894}}.

\bibitem{Bekaert:2012vt}
X.~Bekaert and M.~Grigoriev, {\it {Notes on the ambient approach to boundary
  values of AdS gauge fields}},  \href{http://arXiv.org/abs/1207.3439}{{\tt
  1207.3439}}.

\bibitem{Freedman:1998tz}
D.~Z. Freedman, S.~D. Mathur, A.~Matusis and L.~Rastelli, {\it {Correlation
  functions in the CFT(d) / AdS(d+1) correspondence}},  {\em Nucl.Phys.} {\bf
  B546} (1999) 96--118 [\href{http://arXiv.org/abs/hep-th/9804058}{{\tt
  hep-th/9804058}}].

\bibitem{D'Hoker:1998gd}
E.~D'Hoker and D.~Z. Freedman, {\it {Gauge boson exchange in AdS(d+1)}},  {\em
  Nucl.Phys.} {\bf B544} (1999) 612--632
  [\href{http://arXiv.org/abs/hep-th/9809179}{{\tt hep-th/9809179}}].

\bibitem{Liu:1998ty}
H.~Liu and A.~A. Tseytlin, {\it {On four point functions in the CFT / AdS
  correspondence}},  {\em Phys.Rev.} {\bf D59} (1999) 086002
  [\href{http://arXiv.org/abs/hep-th/9807097}{{\tt hep-th/9807097}}].

\bibitem{D'Hoker:1999jc}
E.~D'Hoker, D.~Z. Freedman, S.~D. Mathur, A.~Matusis and L.~Rastelli, {\it
  {Graviton and gauge boson propagators in AdS(d+1)}},  {\em Nucl.Phys.} {\bf
  B562} (1999) 330--352 [\href{http://arXiv.org/abs/hep-th/9902042}{{\tt
  hep-th/9902042}}].

\bibitem{Fronsdal:1978rb}
C.~Fronsdal, {\it Massless fields with integer spin},  {\em Phys. Rev.} {\bf
  D18} (1978) 3624.

\bibitem{Mikhailov:2002bp}
A.~Mikhailov, {\it {Notes on higher spin symmetries}},
  \href{http://arXiv.org/abs/hep-th/0201019}{{\tt hep-th/0201019}}.

\bibitem{Balasubramanian:1998de}
V.~Balasubramanian, P.~Kraus, A.~E. Lawrence and S.~P. Trivedi, {\it
  {Holographic probes of anti-de Sitter space-times}},  {\em Phys.Rev.} {\bf
  D59} (1999) 104021 [\href{http://arXiv.org/abs/hep-th/9808017}{{\tt
  hep-th/9808017}}].

\bibitem{Iazeolla:2008ix}
C.~Iazeolla and P.~Sundell, {\it {A Fiber Approach to Harmonic Analysis of
  Unfolded Higher- Spin Field Equations}},  {\em JHEP} {\bf 10} (2008) 022
  [\href{http://arXiv.org/abs/0806.1942}{{\tt 0806.1942}}].

\bibitem{Boulanger:2011se}
N.~Boulanger and E.~Skvortsov, {\it {Higher-spin algebras and cubic
  interactions for simple mixed-symmetry fields in AdS spacetime}},  {\em JHEP}
  {\bf 1109} (2011) 063 [\href{http://arXiv.org/abs/1107.5028}{{\tt
  1107.5028}}].

\bibitem{Konshtein:1988yg}
S.~E. Konshtein and M.~A. Vasiliev, {\it Massless representations and
  admissibility condition for higher spin superalgebras},  {\em Nucl. Phys.}
  {\bf B312} (1989) 402.

\bibitem{Stelle:1979aj}
K.~S. Stelle and P.~C. West, {\it Spontaneously broken de sitter symmetry and
  the gravitational holonomy group},  {\em Phys. Rev.} {\bf D21} (1980) 1466.

\bibitem{Vasiliev:1988xc}
M.~A. Vasiliev, {\it Equations of motion of interacting massless fields of all
  spins as a free differential algebra},  {\em Phys. Lett.} {\bf B209} (1988)
  491--497.

\bibitem{Vasiliev:1988sa}
M.~A. Vasiliev, {\it Consistent equations for interacting massless fields of
  all spins in the first order in curvatures},  {\em Annals Phys.} {\bf 190}
  (1989) 59--106.

\bibitem{Vasiliev:2007yc}
M.~A. Vasiliev, {\it On conformal, sl(4,r) and sp(8,r) symmetries of 4d
  massless fields},  \href{http://arXiv.org/abs/arXiv:0707.1085 [hep-th]}{{\tt
  arXiv:0707.1085 [hep-th]}}.

\bibitem{Gelfond:2008td}
O.~Gelfond and M.~Vasiliev, {\it {Sp(8) invariant higher spin theory, twistors
  and geometric BRST formulation of unfolded field equations}},  {\em JHEP}
  {\bf 0912} (2009) 021 [\href{http://arXiv.org/abs/0901.2176}{{\tt
  0901.2176}}].

\bibitem{Vasiliev:2012vf}
M.~A. Vasiliev, {\it {Holography, Unfolding and Higher-Spin Theory}},
  \href{http://arXiv.org/abs/1203.5554}{{\tt 1203.5554}}.

\bibitem{Boulanger:2011dd}
N.~Boulanger and P.~Sundell, {\it {An action principle for Vasiliev's
  four-dimensional higher-spin gravity}},  {\em J.Phys.A} {\bf A44} (2011)
  495402 [\href{http://arXiv.org/abs/1102.2219}{{\tt 1102.2219}}].

\bibitem{Metsaev:1993ap}
R.~R. Metsaev, {\it {Generating function for cubic interaction vertices of
  higher spin fields in any dimension}},  {\em Mod. Phys. Lett.} {\bf A8}
  (1993) 2413--2426.

\bibitem{Metsaev:2005ar}
R.~R. Metsaev, {\it {Cubic interaction vertices for massive and massless higher
  spin fields}},  {\em Nucl. Phys.} {\bf B759} (2006) 147--201
  [\href{http://arXiv.org/abs/hep-th/0512342}{{\tt hep-th/0512342}}].

\bibitem{Zhiboedov:2012bm}
A.~Zhiboedov, {\it {A note on three-point functions of conserved currents}},
  \href{http://arXiv.org/abs/1206.6370}{{\tt 1206.6370}}.

\bibitem{Vasilev:2011xf}
M.~Vasiliev, {\it {Cubic Vertices for Symmetric Higher-Spin Gauge Fields in
  $(A)dS_d$}},  {\em Nucl.Phys.} {\bf B862} (2012) 341--408
  [\href{http://arXiv.org/abs/1108.5921}{{\tt 1108.5921}}].

\end{thebibliography}
\end{document}